\documentclass[prl,aps,twocolumn,showpacs]{revtex4-1}

\usepackage{graphicx}
\usepackage{dcolumn}
\usepackage{bm}
\usepackage{color}
\usepackage{float}
\usepackage{sistyle}
\usepackage{subfigure}  
\usepackage{verbatim}   
\usepackage[citecolor=blue,filecolor=blue,colorlinks=true,urlcolor=blue]{hyperref}
\usepackage{ulem}

\begin{document}

\title{Pressure induced collapse of the charge density wave and Higgs mode visibility in 2H-TaS$_2$} 
\author{Romain Grasset$^{1,}$}\email[]{romain.grasset@univ-paris-diderot.fr}
\author{Yann Gallais$^{1}$}
\author{Alain Sacuto$^{1}$}
\author{Maximilien Cazayous$^{1}$}
\author{Samuel Ma\~nas-Valero$^{2}$}
\author{Eugenio Coronado$^{2}$}
\author{Marie-Aude M\'easson$^{1,3,}$}\email[]{ marie-aude.measson@neel.cnrs.fr}
\affiliation{$^1$Laboratoire Mat\'eriaux et Ph\'enom\`enes Quantiques, UMR 7162 CNRS, Universit\'e Paris Diderot, B$\hat{a}$t. Condorcet 75205 Paris Cedex 13, France\\
$^2$ Universidad de Valencia (ICMol), Catedratico Jos\'e Beltran Martinez, 46980, Paterna, Spain\\
$^3$Institut NEEL CNRS/UGA UPR2940, MCBT, 25 rue des Martyrs BP 166, 38042 Grenoble cedex 9, France
}

\date{\today}

\begin{abstract}
The pressure evolution of the Raman active electronic excitations of the transition metal dichalcogenides 2H-TaS$_2$ is followed through the pressure phase diagram embedding incommensurate charge-density-wave and superconducting states. At high pressure, the charge-density-wave is found to collapse at 8.5~GPa. In the coexisting charge-density-wave and superconducting orders, we unravel a strong in-gap superconducting mode, attributed to a Higgs mode, coexisting with the expected incoherent Cooper-pair breaking signature. The latter remains in the pure superconducting state reached above  8.5~GPa. Our report constitutes the first observation of such Raman active Higgs mode since the longstanding unique case 2H-NbSe$_2$.

\end{abstract}

\maketitle

Symmetry breaking across an electronic phase transition always occur along with the emergence of new collective excitations, including oscillations of the amplitude of the order parameter. In charge-density-wave (CDW) systems, translational symmetry breaking gives rise to the amplitudon \cite{lee_conductivity_1974,gruner_review}. Similarly, in superconductors U(1) rotational symmetry breaking gives rise to oscillations of the amplitude of the order parameter, also called Higgs mode because of its analogy to the one found in high-energy field theories \cite{littlewood_amplitude_1982}. The study of these collective modes and their interaction is of great interest for the study and control of intertwined electronic orders. In the case of co-existing CDW and superconducting (SC) orders, recent studies \cite{sentef_theory_2017,PhysRevB.93.195139,PhysRevB.91.174502,Fausti189} have shown that controlling order parameters dynamics of these co-existing orders using light pulses can induce an enhancement of the superconducting critical temperature. In the context of high $T_c$ cuprates  co-existing SC and CDW orders have also attracted great interest recently since they could lead to non-uniform SC state, a pair density wave (PDW),  with a distinctive order parameter dynamics \cite{hamidian_detection_2016, PhysRevB.95.214502}.

Raman spectroscopy is a well-known tool for the observation of excitations in materials. It has been extensively used to study amplitudons in various systems exhibiting a CDW order \cite{tsang_raman_1976,sugai_phason_2006,travaglini_CDW_1983}. However the observation of a Higgs mode in a superconducting (SC) state remains elusive as it only weakly coupled to spectroscopic probes \cite{cea_thg_prb16} and remains short-lived because of the quasiparticle continuum developing at 2$\Delta$ \citep{volkov73,kulik81,cea_nature_2014,cea_prl15}. Nevertheless the 
detection of the Higgs mode has been recently reported using strong 
THz pulses in conventional \cite{shimano_prl13,shimano_science14} and unconventional SC 
\cite{PhysRevLett.120.117001}, and also by conventional infrared (IR) spectroscopy in 
disordered SC \cite{frydman_natphys15}. However nature of the measured mode and the conditions of its observability are still 
under debate \cite{pekker_amplitudehiggs_2015,cea_prl15,cea_thg_prb16,armitage_prb16,tsuji_prb16,PhysRevB.96.020505,PhysRevB.97.094516}. Early on a possible observation of the Higgs mode was also reported in the transition metal dichalcogenide (TMDC) 2H-NbSe$_2$ using Raman spectroscopy \cite{sooryakumar_raman_1980,measson_amplitude_2014,grasset_prb_2018} where the Higgs mode could be visible thanks to the coupling to the CDW amplitudon  \cite{littlewood_amplitude_1982,browne_prb83,cea_nature_2014}. At present 2H-NbSe$_2$ remains a unique case and other observation of Higgs modes in CDW superconductors are desirable to assess how generic is the coupling between the Higgs and the amplitudon.

Although in high-T$_c$ cuprates a coexisting CDW and SC phase has recently been detected \cite{ghiringhelli_long-range_2012,da_silva_neto_ubiquitous_2014,comin_charge_2014,wu_magnetic-field-induced_2011,tranquada_evidence_1995,lake_antiferromagnetic_2002}, the absence of a long-range CDW order or the d-wave nature of the SC gap could explain that no Higgs mode has yet been identified in the Raman spectra. 
On the other hand, the family of the TMDCs contains few systems where SC and CDW orders coexist\cite{sipos_mott_2008,kusmartseva_pressure_2009,bhoi_interplay_2016}. Generally low superconducting critical temperatures prevents any Raman spectroscopy study of the SC state. In 2H-TaS$_2$, an incommensurate CDW develops below 77~K followed by a superconducting state below T$_c$=1~K. Recent reports \cite{freitas_strong_2016} have shown a dramatic increase in T$_c$ with pressure, up to 8.5~K at 10~GPa. Then an observation of the coexisting superconducting and CDW states, and of the Higgs mode, becomes accessible using Raman scattering.

In this Letter we map out the CDW phase diagram of 2H-TaS$_2$ by following the CDW excitations and gap under high pressure and find that the CDW completely collapses at 8.5~GPa. In the low temperature SC state, we unravel a low energy in-gap collective mode which is attributed to a Higgs mode and whose interplay with the CDW mode points to a similar mechanism of observability as in 2H-NbSe$_{2}$. Beside, in 2H-TaS$_2$, this in-gap mode coexists with the usual incoherent Cooper-pair breaking peak at 2$\Delta$ clearly differentiating both excitations, and demonstrating that the Higgs mode is a well-defined collective mode located below the continuum of quasiparticle excitations in the SC+CDW state.

Crystals of 2H-TaS$_2$ were grown by chemical vapour transport from the pre-synthetic material, using iodine as a transport agent, as already reported \cite{navarro-moratalla_enhanced_2016}. Composition and phase purity were confirmed by powder X-ray diffraction, inductively coupled plasma spectrometry and elemental analysis (see Supplemental material for further information). We have performed Raman spectroscopy measurements on single crystals of bulk 2H-TaS$_2$ under hydrostatic pressure in a membrane diamond anvil cell using Helium as a pressure-transmitting medium. We have tracked low energy excitations down to 7~cm$^{-1}$ under extreme conditions \cite{buhot_driving_2015,grasset_prb_2018}, down to 3~K and up to 9.5~GPa. Superconductivity was accessed by performing measurements at low laser power of 0.2~mW. We have followed simultaneously the phonons, the charge-density-wave modes and the superconducting excitations across the Pressure-Temperature phase diagram.

\begin{figure}[!h]
\centering
\includegraphics[width=0.95\linewidth]{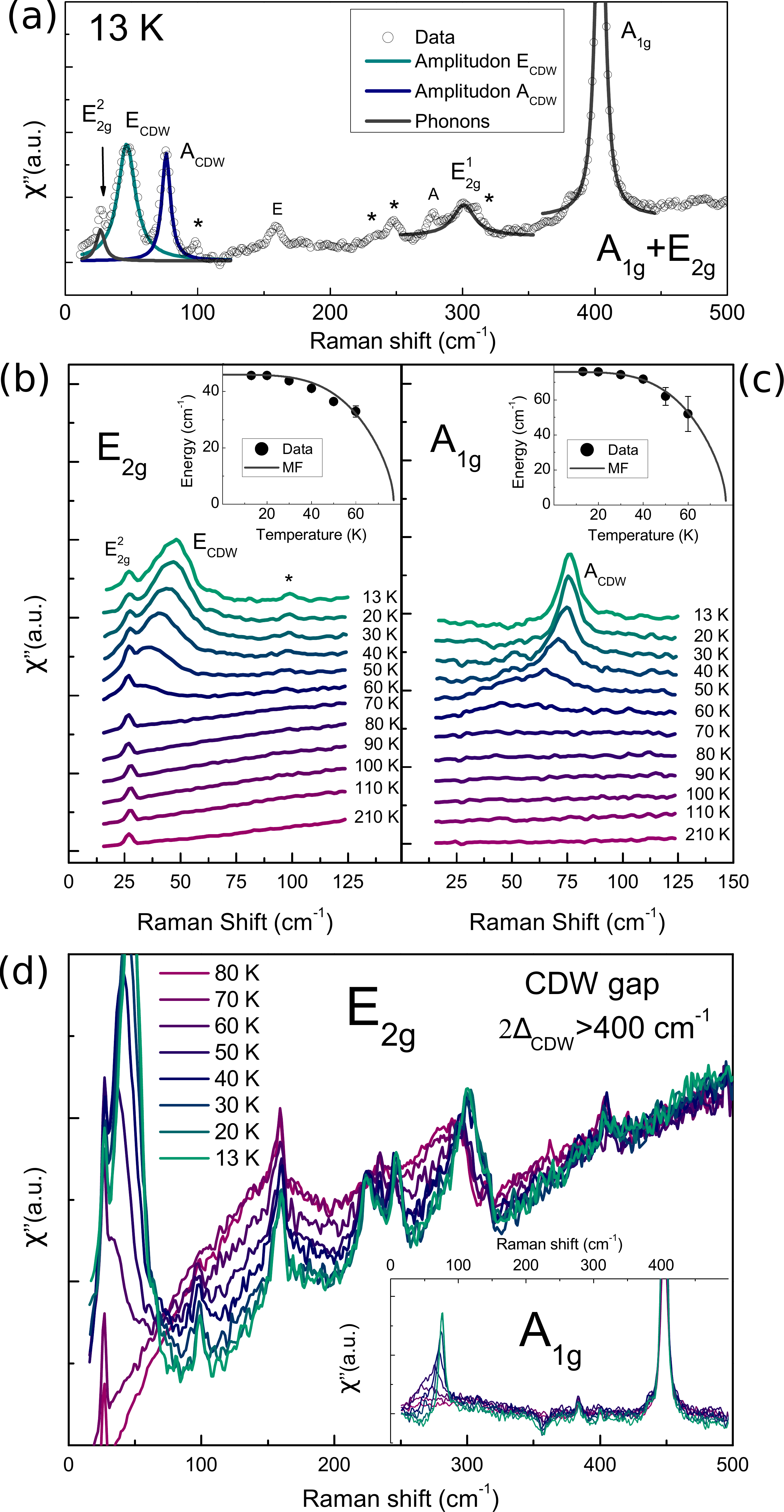} 
\caption{a) Raman spectra of 2H-TaS$_2$ at ambient pressure at 13~K in the ICDW state and for the A$_{1g}$+E$_{2g}$ symmetry. We observe three phonons (E$^2_{2g}$, E$^1_{2g}$ and A$_{1g}$) and two CDW modes labelled according to there symmetry (E$_{CDW}$ and A$_{CDW}$). Additional less intense CDW modes (denoted *) are also visible. (b,c) Temperature dependence of the amplitudons in the pure symmetries. Insets: Energy of the amplitudons as a function of temperature. The black line corresponds to a fit using a mean-field calculation. d) Raman spectra at various temperatures in the E$_{2g}$ symmetry where a CDW gap opens. Inset: Raman spectra in the A$_{1g}$ symmetry.} 
\label{ambient}
\end{figure}

In Fig.~\ref{ambient}a) we show the Raman response from 2H-TaS$_2$ at ambient pressure at 13~K. Three regular phonons are measured: E$_{2g}^2$ (26.1~cm$^{-1}$), E$_{2g}^1$ (300.3~cm$^{-1}$) and A$_{1g}$ (404.0~cm$^{-1}$). The incommensurate charge-density-wave (ICDW) manifests it-self with amplitudons, which correspond to a soft-phonon coupled to the electronic density at Q$_{CDW}$ and dressed by the amplitude fluctuations of the CDW order parameter\cite{gruner_review,lee_conductivity_1974}. Contrary to previous works \cite{sugai_studies_1981} where only one amplitudon could be readily tracked, we report two well-defined amplitudons, one in each symmetry and labelled accordingly: E$_{CDW}$ (46.5~cm$^{-1}$) and A$_{CDW}$ (75.8~cm$^{-1}$). As shown Fig.~\ref{ambient}b), the CDW amplitudons' behaviors at ambient pressure is typical to this kind of excitations: both loose intensity, enlarge and soften towards zero energy with increasing temperature toward the transition at 77~K. In the inset of Fig.~\ref{ambient} b,c) the amplitudon energy are displayed as function of temperature. They are well fitted using a mean-field like temperature dependence  \footnote{$(1-\frac{x^4}{3})\sqrt{1-x^4}$ where $x=\frac{T}{T_{CDW}}$ taken from Ref.\cite{Benfatto2000} and which reproduces
a mean-field-like behaviour near both $T=0$ and $T_{CDW}$}\label{MF}. This typical order parameter-like behavior is also observed for the CDW amplitudons of 2H-NbSe$_2$ \cite{tsang_raman_1976}.


In addition, only in the ICDW state, we observe multiple weak peaks, labelled (*), which most probably correspond to regular phonons folded to the zone center of the Brillouin zone due to the establishment of the CDW state. While keeping the same energy, the folded CDW phonons (*) smoothly loose intensity and disappear at T$_{ICDW}$.

Beside, as presented Fig.~\ref{ambient}d), a depletion  of the electronic background develops in the E$_{2g}$ symmetry below T$_{ICDW}$. This loss of spectral weight at low energy is attributed to the opening of a gap in the CDW state, similarly to what has been observed in rare-earth tellurides prototopical CDW systems \cite{eiter_alternative_2013,ralevic_charge_2016}. The gap extends up to at least $\Delta$=400 cm$^{-1}$. On contrary, no such a gap is measured in the A$_{1g}$ symmetry (See Inset of Fig.~\ref{ambient}d), either because of the screening of the electronic response in this channel due to the Coulomb effect or because of a significant anisotropy of the CDW gap. Interestingly, the CDW gap in 2H-TaS$_2$ appears clearly by Raman scattering whereas it remains elusive in 2H-NbSe$_2$. This is probably due to the fact that the CDW gap is open on larger parts of the Fermi surface in 2H-TaS$_2$, in good agreement with recent ARPES results \cite{zhao_orbital_2017}. \\

Two peaks are also detected at $\sim$160~cm$^{-1}$ (E) and $\sim$270~cm$^{-1}$ (A) in the E$_{2g}$ and A$_{1g}$ symmetries, respectively. They are measured already at 300~K in the normal state and persist up to the highest pressure. They are not associated to the CDW state and may arise from IR phonons activated by disorder. 
\\

\begin{figure}[!h]
\includegraphics[width=\linewidth]{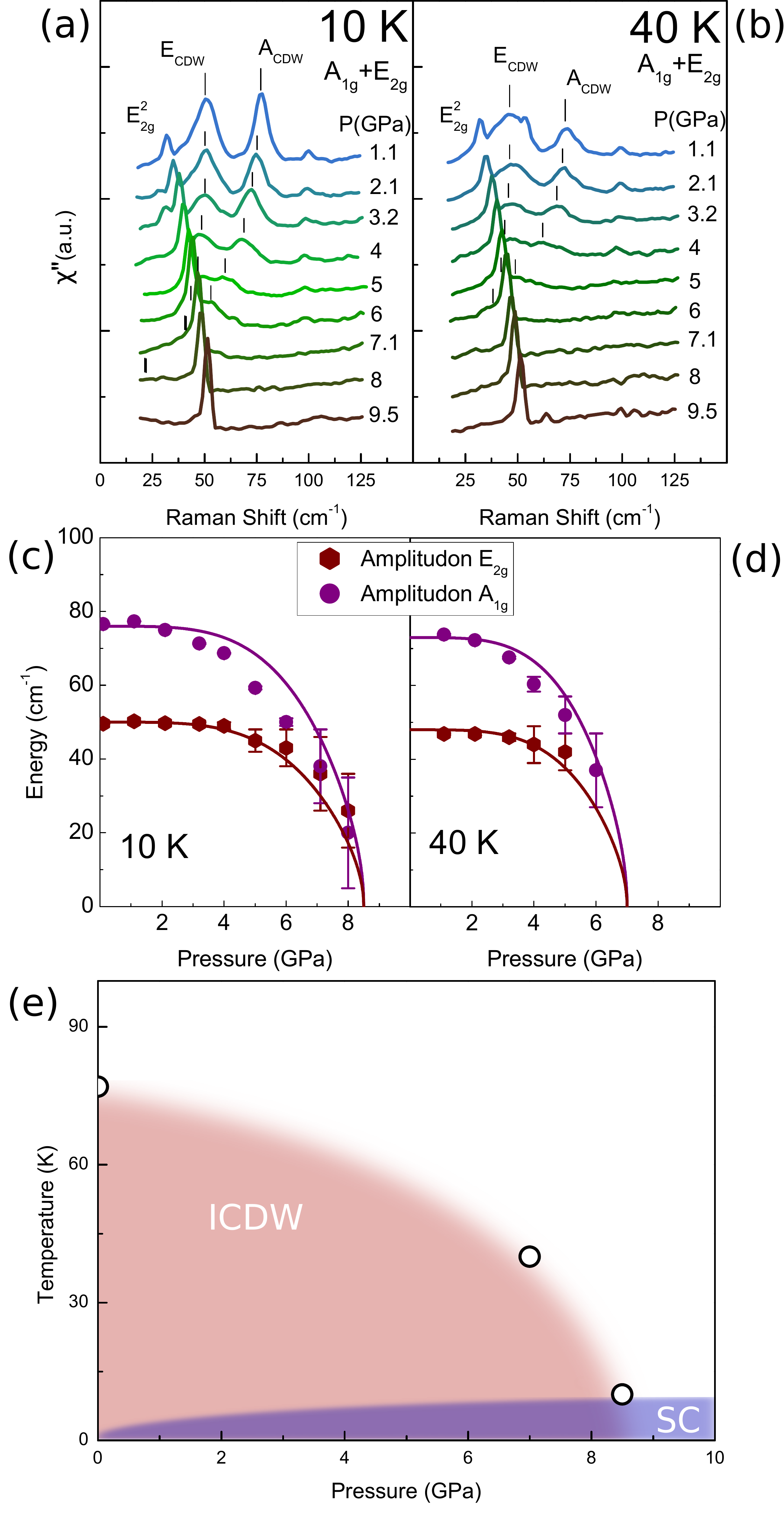} 
\caption{(a,b) Raman spectra of 2H-TaS$_2$ in the A$_{1g}$+E$_{2g}$ symmetry at 10~K (a) and 40~K (b) for various pressures ranging from 1.1~GPa to 9.5~GPa. (c,d) Pressure evolution of the energy of the CDW amplitudons at 10~K (c) and 40~K (d). The A$_{1g}$ and E$_{2g}$ modes soften with pressure towards zero energy (solid lines are guides for the eye). (e) Phase diagram (T,P) of 2H-TaS$_2$.}
\label{cdw}
\end{figure}

Fig.\ref{cdw} a,b) show the pressure evolution of the Raman spectra of 2H-TaS$_2$ in the A$_{1g}$+E$_{2g}$ symmetry at 10~K and 40~K, respectively. While the E$^2_{2g}$ phonon harden with pressure, its width remains stable showing a good hydrostaticity in the pressure chamber. Both A$_{1g}$ and E$_{2g}$ amplitudons soften and enlarge as the pressure is increased up to a complete collapse between 8 and 9.5~GPa at 10~K. (and between 6 and 7.1~GPa at 40~K).


As shown Fig. \ref{cdw} c,d), the CDW modes present an order parameter-like behavior similar to the temperature dependence. Using equation \footnote{Equation of Ref.[39] using $x=\frac{P}{P_c}$} to follow the evolution of the mode energies, the collapse of the CDW occurs at 8.5~GPa for 10~K and  7~GPa at 40~K.

Further evidences of the CDW collapse above 8.5~GPa are found in the folded CDW phonons (*) and the CDW gap (See Supplemental material), which are no longer measured above this critical pressure. All three manifestations of the CDW in 2H-TaS$_2$ are consistent and point toward a complete collapse of the CDW at 8.5~GPa.

Hence we draw a new phase diagram for the CDW in 2H-TaS$_2$ using Raman spectroscopy, as depicted in Fig.~\ref{cdw}e). Notable differences are obtained with previous results from transport measurement \cite{freitas_strong_2016} where  signature of the ICDW were reported up to at least 17~GPa. This discrepancy might result from the presence of a 
 pseudogap as reported by ARPES measurements \cite{zhao_orbital_2017} at ambient pressure above $T_{CDW}$ and which may survive above P$_c$ while being detected by transport measurements. Alternatively, application of high pressure could induce a loss of a long-range CDW order and a softening of the amplitudon energy \cite{chen_electronic_1997,sugai_carrier-density-dependent_2003}, while a short-range CDW order may leave a broad signature detected by transport measurements.

\begin{figure}[!ht]
\includegraphics[width=\linewidth]{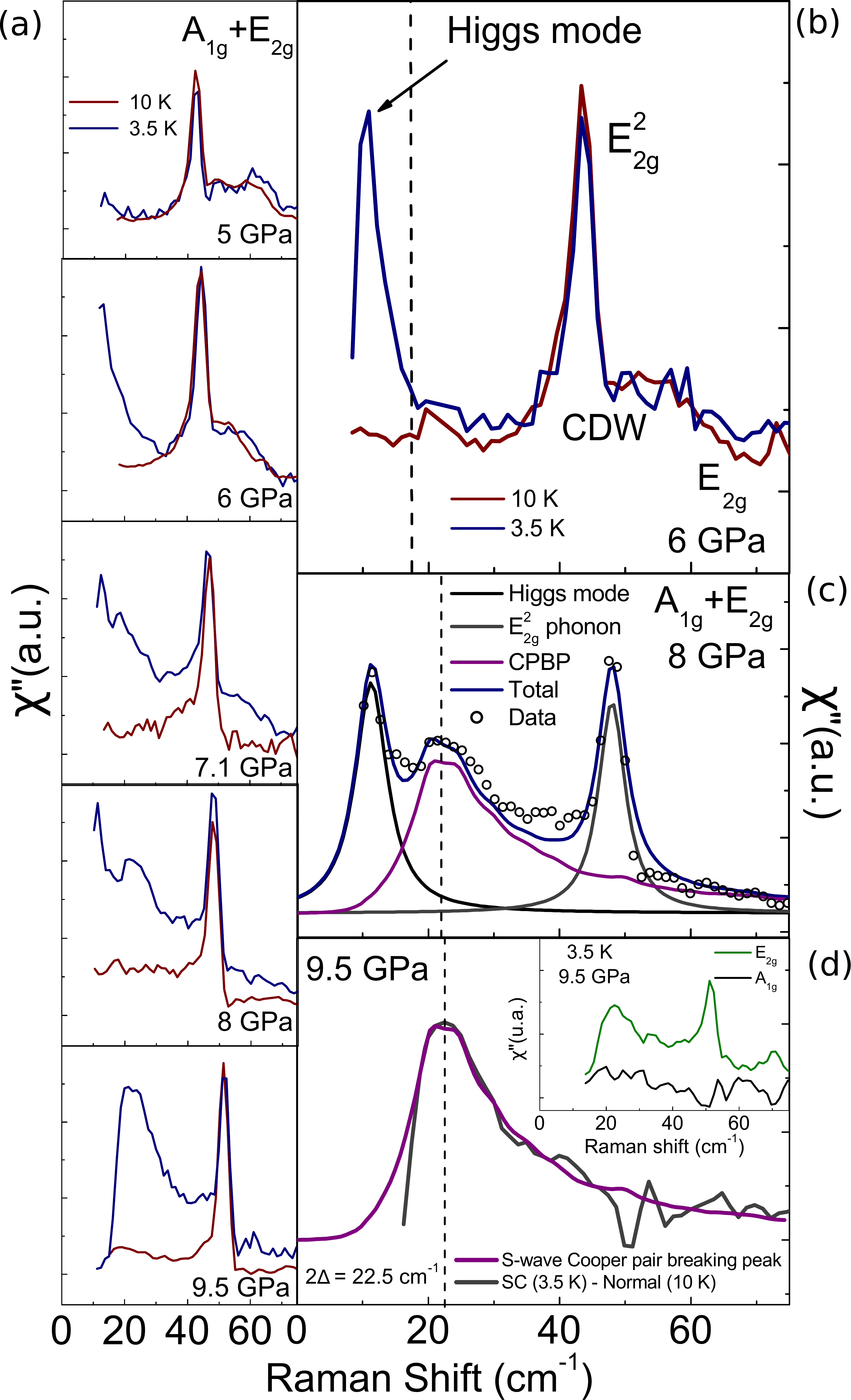}
\caption{(a) Raman spectra of 2H-TaS$_2$ in the A$_{1g}$+E$_{2g}$ symmetry at 3.5~K and 10~K for various pressure ranging from 5 to 9.5~GPa. (b) Low energy Raman spectra in the E$_{2g}$ symmetry at 6~GPa below T$_c$. The in-gap mode is at $\sim$ 10~cm$^{-1}$, below 2$\Delta$ (dotted line), similarly to what is observed in 2H-NbSe$_2$. (c) Fit of the various features observed in the SC state at 8~GPa in the A$_{1g}$+E$_{2g}$ symmetry: a sharp in-gap Higgs mode, a incoherent Cooper-pair breaking peak (CPBP) and the E$_{2g}^2$ phonon. (d) Subtraction of the normal state response (10~K) from the superconducting state response (3.5~K) at 9.5~GPa in the A$_{1g}$+E$_{2g}$ symmetry. The SC signature fits a Cooper-pair breaking peak for a BCS response in an s-wave superconductor with a gap 2$\Delta$=22.5~cm$^{-1}$ corresponding to a T$_c$ of 8.55~K (purple line). Inset: Raman spectra for the pure A$_{1g}$ (black curve) and E$_{2g}$ (green curve) symmetries. The Cooper-pair breaking peak appears only in the E$_{2g}$ symmetry.}
\label{higgs}
\end{figure}

We now turn to the study of the superconducting state, reached above 5~GPa by minimizing the laser heating. While entering it, new features develop above 6~GPa ($T_c>$6.5~K) (See Fig.~\ref{higgs} (a)). At 6~GPa, a low energy superconducting excitation starts to develop. It is visible only in the SC state vanishing completely above T$_c$. Its shape and position is confirmed by a second pressure run spectra reaching lower energies ($\sim$8~cm$^{-1}$). It is a narrow and intense in-gap mode (below 2$\Delta$, See Supplemental material) and it is present, at least, in the E$_{2g}$ symmetry (See Fig.~\ref{higgs} (b)). 

While increasing further the applied pressure, an additional feature develops at 2$\Delta$ (See Supplemental material). It consists in a gap opening below $\sim$20~cm$^{-1}$ and an asymmetric peak above (See Fig. \ref{higgs} c). This structure, sometimes observed in simple superconductors \cite{klein_1984} and here observed up to 9.5~GPa in the pure superconducting state, is the expected incoherent Cooper-pair breaking peak (CPBP). At the highest measured pressure (9.5~GPa), the SC transition has already reached its maximum of 8.5~K\cite{freitas_strong_2016}. By calculating the theoretical Raman response (purple line in Fig. \ref{higgs} (d)) of an s-wave superconductor, a gap $2\Delta$ of 22.5~cm$^{-1}$ is obtained. This corresponds to a T$_c$ of 8.85~K, using the standard weak-coupling BCS ratio, in good agreement with the T$_c$ measured by transport measurements\cite{freitas_strong_2016}. We note that it is only visible in the E$_{2g}$ symmetry (See inset of Fig.~\ref{higgs} d)) likely due to Coulomb screening in the A$_{1g}$ channel \cite{devereaux_electronic_1995,devereaux_inelastic_2007,cea_raman_prb16}.
Above P$_c$, at 9.5~GPa, the in-gap mode disappears with the collapse of the CDW order, thus mimicking the behavior of the in-gap mode measured in 2H-NbSe$_2$ \cite{sooryakumar_raman_1980,measson_amplitude_2014,grasset_prb_2018}. Both modes in these two brothers compounds certainly share the same nature. Up to now, the most explored hypothesis, supported by theoretical calculations and investigation under high pressure, of the nature of the in-gap modes in 2H-NbSe$_2$, and so of this pressure-induced one in 2H-TaS$_2$, is its assignment to the amplitude 'Higgs' mode \cite{littlewood_amplitude_1982,browne_prb83,cea_nature_2014}, the analogous of the Higgs boson in superconductors. 

Here, the observation of both superconducting features in the SC+CDW state, the in-gap mode and the incoherent Cooper-pair breaking peak  at well separated energies is crucial. The evolution of the Cooper-pair breaking peak through the whole pressure phase diagram from the coexisting CDW+SC state to the pure SC state is gradual: the energy follows the evolution of T$_c$ and the spectral weight continuously increases as the SC gap takes over parts of the Fermi surface previously gapped by the CDW. In particular we do not observe any dramatic effect of the collapse of the CDW order on the incoherent CPBP. By contrast the in-gap mode intensity abruptly collapses in the pure SC state, as expected for the SC Higgs mode which couples to the Raman probe only via the CDW order. These observations rule out the interpretation of the in-gap mode as a Cooper-pair breaking peak affected by the opening of the CDW gap. They further demonstrate that the Higgs mode is a collective mode located below the continuum of quasiparticle excitations in the coexistence state consistently with theoretical work \cite{littlewood_amplitude_1982,cea_nature_2014}.

The proposed mechanism of observability of the Higgs mode in the presence of CDW amplitudons \cite{littlewood_amplitude_1982,browne_prb83,cea_nature_2014} has been shown to be consistent with the pressure dependence of the electronic excitations in NbSe$_2$ \cite{grasset_prb_2018}. The present data suggest this mechanism is also at play in 2H-TaS$_2$. This would then imply that in this compound, even if the Fermi surface is significantly gapped by the CDW \cite{zhao_orbital_2017}, the superconducting and CDW gaps must overlap on some parts of the Fermi surface.




In conclusion, from Raman scattering, we draw a Pressure-Temperature phase diagram of 2H-TaS$_2$. The incommensurate charge-density-wave completely collapses at about 8.5~GPa. In the coexisting charge-density-wave and superconducting state, an in-gap superconducting mode, interpreted as a Higgs mode, is reported.  This constitutes the first observation of such Raman active Higgs mode in condensed matter systems since the unique case of 2H-NbSe$_2$. It has been clearly differentiated from the usual incoherent Cooper-pair breaking peak 
 which survives in the pressure-induced pure superconducting state and our observations are consistent with the mechanism of observability of the Higgs mode for which an overlap of the charge-density-wave and superconducting gaps on the Fermi surface is necessary. This work paves the way for the systematic search of Higgs mode in superconductors and the study of its visibility while diversifying the coexisting orders.

\section*{Acknowledgments}
We thank gratefully L. Benfatto, C. Varma, H. Suderow and I. Paul for fruitful discussions and G. Lemarchand
and A. Polian for technical support.
This work was supported by the Labex SEAM (Grant No. ANR-11-IDEX-0005-02)and by the French Agence Nationale de la Recherche (ANR PRINCESS, Grant No. ANR-11-BS04-002 and ANR SEO-HiggS2, Grant No. ANR-16-CE30-0014). E.C. and S.M.-V. acknowledge the financial support from the European Research Council (Advance  Grant Mol-2D ref. 788222), the Spanish MINECO (Excellence Unit “María de Maeztu” MDM- 2015-0538), and the Generalitat Valenciana (PROMETEO/2017/066). S.M.-V. thanks the Spanish MECD for a F.P.U. doctoral fellowship (FPU014/04407).


\begin{thebibliography}{55}%
\makeatletter
\providecommand \@ifxundefined [1]{%
 \@ifx{#1\undefined}
}%
\providecommand \@ifnum [1]{%
 \ifnum #1\expandafter \@firstoftwo
 \else \expandafter \@secondoftwo
 \fi
}%
\providecommand \@ifx [1]{%
 \ifx #1\expandafter \@firstoftwo
 \else \expandafter \@secondoftwo
 \fi
}%
\providecommand \natexlab [1]{#1}%
\providecommand \enquote  [1]{``#1''}%
\providecommand \bibnamefont  [1]{#1}%
\providecommand \bibfnamefont [1]{#1}%
\providecommand \citenamefont [1]{#1}%
\providecommand \href@noop [0]{\@secondoftwo}%
\providecommand \href [0]{\begingroup \@sanitize@url \@href}%
\providecommand \@href[1]{\@@startlink{#1}\@@href}%
\providecommand \@@href[1]{\endgroup#1\@@endlink}%
\providecommand \@sanitize@url [0]{\catcode `\\12\catcode `\$12\catcode
  `\&12\catcode `\#12\catcode `\^12\catcode `\_12\catcode `\%12\relax}%
\providecommand \@@startlink[1]{}%
\providecommand \@@endlink[0]{}%
\providecommand \url  [0]{\begingroup\@sanitize@url \@url }%
\providecommand \@url [1]{\endgroup\@href {#1}{\urlprefix }}%
\providecommand \urlprefix  [0]{URL }%
\providecommand \Eprint [0]{\href }%
\providecommand \doibase [0]{http://dx.doi.org/}%
\providecommand \selectlanguage [0]{\@gobble}%
\providecommand \bibinfo  [0]{\@secondoftwo}%
\providecommand \bibfield  [0]{\@secondoftwo}%
\providecommand \translation [1]{[#1]}%
\providecommand \BibitemOpen [0]{}%
\providecommand \bibitemStop [0]{}%
\providecommand \bibitemNoStop [0]{.\EOS\space}%
\providecommand \EOS [0]{\spacefactor3000\relax}%
\providecommand \BibitemShut  [1]{\csname bibitem#1\endcsname}%
\let\auto@bib@innerbib\@empty
\bibitem [{\citenamefont {Lee}\ \emph {et~al.}(1974)\citenamefont {Lee},
  \citenamefont {Rice},\ and\ \citenamefont
  {Anderson}}]{lee_conductivity_1974}%
  \BibitemOpen
  \bibfield  {author} {\bibinfo {author} {\bibfnamefont {P.~A.}\ \bibnamefont
  {Lee}}, \bibinfo {author} {\bibfnamefont {T.~M.}\ \bibnamefont {Rice}}, \
  and\ \bibinfo {author} {\bibfnamefont {P.~W.}\ \bibnamefont {Anderson}},\
  }\href {\doibase 10.1016/0038-1098(74)90868-0} {\bibfield  {journal}
  {\bibinfo  {journal} {Solid State Commun.}\ }\textbf {\bibinfo {volume}
  {14}},\ \bibinfo {pages} {703} (\bibinfo {year} {1974})}\BibitemShut
  {NoStop}%
\bibitem [{\citenamefont {Gr\"uner}(1988)}]{gruner_review}%
  \BibitemOpen
  \bibfield  {author} {\bibinfo {author} {\bibfnamefont {G.}~\bibnamefont
  {Gr\"uner}},\ }\href {\doibase 10.1103/RevModPhys.60.1129} {\bibfield
  {journal} {\bibinfo  {journal} {Rev. Mod. Phys.}\ }\textbf {\bibinfo {volume}
  {60}},\ \bibinfo {pages} {1129} (\bibinfo {year} {1988})}\BibitemShut
  {NoStop}%
\bibitem [{\citenamefont {Littlewood}\ and\ \citenamefont
  {Varma}(1982)}]{littlewood_amplitude_1982}%
  \BibitemOpen
  \bibfield  {author} {\bibinfo {author} {\bibfnamefont {P.}~\bibnamefont
  {Littlewood}}\ and\ \bibinfo {author} {\bibfnamefont {C.}~\bibnamefont
  {Varma}},\ }\href {\doibase 10.1103/PhysRevB.26.4883} {\bibfield  {journal}
  {\bibinfo  {journal} {Phys. Rev. B}\ }\textbf {\bibinfo {volume} {26}},\
  \bibinfo {pages} {4883} (\bibinfo {year} {1982})}\BibitemShut {NoStop}%
\bibitem [{\citenamefont {Sentef}\ \emph {et~al.}(2017)\citenamefont {Sentef},
  \citenamefont {Tokuno}, \citenamefont {Georges},\ and\ \citenamefont
  {Kollath}}]{sentef_theory_2017}%
  \BibitemOpen
  \bibfield  {author} {\bibinfo {author} {\bibfnamefont {M.~A.}\ \bibnamefont
  {Sentef}}, \bibinfo {author} {\bibfnamefont {A.}~\bibnamefont {Tokuno}},
  \bibinfo {author} {\bibfnamefont {A.}~\bibnamefont {Georges}}, \ and\
  \bibinfo {author} {\bibfnamefont {C.}~\bibnamefont {Kollath}},\ }\href
  {\doibase 10.1103/PhysRevLett.118.087002} {\bibfield  {journal} {\bibinfo
  {journal} {Phys. Rev. Lett.}\ }\textbf {\bibinfo {volume} {118}},\ \bibinfo
  {pages} {087002} (\bibinfo {year} {2017})}\BibitemShut {NoStop}%
\bibitem [{\citenamefont {Patel}\ and\ \citenamefont
  {Eberlein}(2016)}]{PhysRevB.93.195139}%
  \BibitemOpen
  \bibfield  {author} {\bibinfo {author} {\bibfnamefont {A.~A.}\ \bibnamefont
  {Patel}}\ and\ \bibinfo {author} {\bibfnamefont {A.}~\bibnamefont
  {Eberlein}},\ }\href {\doibase 10.1103/PhysRevB.93.195139} {\bibfield
  {journal} {\bibinfo  {journal} {Phys. Rev. B}\ }\textbf {\bibinfo {volume}
  {93}},\ \bibinfo {pages} {195139} (\bibinfo {year} {2016})}\BibitemShut
  {NoStop}%
\bibitem [{\citenamefont {Casandruc}\ \emph {et~al.}(2015)\citenamefont
  {Casandruc}, \citenamefont {Nicoletti}, \citenamefont {Rajasekaran},
  \citenamefont {Laplace}, \citenamefont {Khanna}, \citenamefont {Gu},
  \citenamefont {Hill},\ and\ \citenamefont {Cavalleri}}]{PhysRevB.91.174502}%
  \BibitemOpen
  \bibfield  {author} {\bibinfo {author} {\bibfnamefont {E.}~\bibnamefont
  {Casandruc}}, \bibinfo {author} {\bibfnamefont {D.}~\bibnamefont
  {Nicoletti}}, \bibinfo {author} {\bibfnamefont {S.}~\bibnamefont
  {Rajasekaran}}, \bibinfo {author} {\bibfnamefont {Y.}~\bibnamefont
  {Laplace}}, \bibinfo {author} {\bibfnamefont {V.}~\bibnamefont {Khanna}},
  \bibinfo {author} {\bibfnamefont {G.~D.}\ \bibnamefont {Gu}}, \bibinfo
  {author} {\bibfnamefont {J.~P.}\ \bibnamefont {Hill}}, \ and\ \bibinfo
  {author} {\bibfnamefont {A.}~\bibnamefont {Cavalleri}},\ }\href {\doibase
  10.1103/PhysRevB.91.174502} {\bibfield  {journal} {\bibinfo  {journal} {Phys.
  Rev. B}\ }\textbf {\bibinfo {volume} {91}},\ \bibinfo {pages} {174502}
  (\bibinfo {year} {2015})}\BibitemShut {NoStop}%
\bibitem [{\citenamefont {Fausti}\ \emph {et~al.}(2011)\citenamefont {Fausti},
  \citenamefont {Tobey}, \citenamefont {Dean}, \citenamefont {Kaiser},
  \citenamefont {Dienst}, \citenamefont {Hoffmann}, \citenamefont {Pyon},
  \citenamefont {Takayama}, \citenamefont {Takagi},\ and\ \citenamefont
  {Cavalleri}}]{Fausti189}%
  \BibitemOpen
  \bibfield  {author} {\bibinfo {author} {\bibfnamefont {D.}~\bibnamefont
  {Fausti}}, \bibinfo {author} {\bibfnamefont {R.~I.}\ \bibnamefont {Tobey}},
  \bibinfo {author} {\bibfnamefont {N.}~\bibnamefont {Dean}}, \bibinfo {author}
  {\bibfnamefont {S.}~\bibnamefont {Kaiser}}, \bibinfo {author} {\bibfnamefont
  {A.}~\bibnamefont {Dienst}}, \bibinfo {author} {\bibfnamefont {M.~C.}\
  \bibnamefont {Hoffmann}}, \bibinfo {author} {\bibfnamefont {S.}~\bibnamefont
  {Pyon}}, \bibinfo {author} {\bibfnamefont {T.}~\bibnamefont {Takayama}},
  \bibinfo {author} {\bibfnamefont {H.}~\bibnamefont {Takagi}}, \ and\ \bibinfo
  {author} {\bibfnamefont {A.}~\bibnamefont {Cavalleri}},\ }\href {\doibase
  10.1126/science.1197294} {\bibfield  {journal} {\bibinfo  {journal}
  {Science}\ }\textbf {\bibinfo {volume} {331}},\ \bibinfo {pages} {189}
  (\bibinfo {year} {2011})}\BibitemShut {NoStop}%
\bibitem [{\citenamefont {Hamidian}\ \emph {et~al.}(2016)\citenamefont
  {Hamidian}, \citenamefont {Edkins}, \citenamefont {Joo}, \citenamefont
  {Kostin}, \citenamefont {Eisaki}, \citenamefont {Uchida}, \citenamefont
  {Lawler}, \citenamefont {Kim}, \citenamefont {Mackenzie}, \citenamefont
  {Fujita}, \citenamefont {Lee},\ and\ \citenamefont
  {Davis}}]{hamidian_detection_2016}%
  \BibitemOpen
  \bibfield  {author} {\bibinfo {author} {\bibfnamefont {M.~H.}\ \bibnamefont
  {Hamidian}}, \bibinfo {author} {\bibfnamefont {S.~D.}\ \bibnamefont
  {Edkins}}, \bibinfo {author} {\bibfnamefont {S.~H.}\ \bibnamefont {Joo}},
  \bibinfo {author} {\bibfnamefont {A.}~\bibnamefont {Kostin}}, \bibinfo
  {author} {\bibfnamefont {H.}~\bibnamefont {Eisaki}}, \bibinfo {author}
  {\bibfnamefont {S.}~\bibnamefont {Uchida}}, \bibinfo {author} {\bibfnamefont
  {M.~J.}\ \bibnamefont {Lawler}}, \bibinfo {author} {\bibfnamefont {E.-A.}\
  \bibnamefont {Kim}}, \bibinfo {author} {\bibfnamefont {A.~P.}\ \bibnamefont
  {Mackenzie}}, \bibinfo {author} {\bibfnamefont {K.}~\bibnamefont {Fujita}},
  \bibinfo {author} {\bibfnamefont {J.}~\bibnamefont {Lee}}, \ and\ \bibinfo
  {author} {\bibfnamefont {J.~C.~S.}\ \bibnamefont {Davis}},\ }\href
  {http://dx.doi.org/10.1038/nature17411} {\bibfield  {journal} {\bibinfo
  {journal} {Nature}\ }\textbf {\bibinfo {volume} {532}},\ \bibinfo {pages}
  {343} (\bibinfo {year} {2016})}\BibitemShut {NoStop}%
\bibitem [{\citenamefont {Soto-Garrido}\ \emph {et~al.}(2017)\citenamefont
  {Soto-Garrido}, \citenamefont {Wang}, \citenamefont {Fradkin},\ and\
  \citenamefont {Cooper}}]{PhysRevB.95.214502}%
  \BibitemOpen
  \bibfield  {author} {\bibinfo {author} {\bibfnamefont {R.}~\bibnamefont
  {Soto-Garrido}}, \bibinfo {author} {\bibfnamefont {Y.}~\bibnamefont {Wang}},
  \bibinfo {author} {\bibfnamefont {E.}~\bibnamefont {Fradkin}}, \ and\
  \bibinfo {author} {\bibfnamefont {S.~L.}\ \bibnamefont {Cooper}},\ }\href
  {\doibase 10.1103/PhysRevB.95.214502} {\bibfield  {journal} {\bibinfo
  {journal} {Phys. Rev. B}\ }\textbf {\bibinfo {volume} {95}},\ \bibinfo
  {pages} {214502} (\bibinfo {year} {2017})}\BibitemShut {NoStop}%
\bibitem [{\citenamefont {Tsang}\ \emph {et~al.}(1976)\citenamefont {Tsang},
  \citenamefont {Smith},\ and\ \citenamefont {Shafer}}]{tsang_raman_1976}%
  \BibitemOpen
  \bibfield  {author} {\bibinfo {author} {\bibfnamefont {J.}~\bibnamefont
  {Tsang}}, \bibinfo {author} {\bibfnamefont {J.}~\bibnamefont {Smith}}, \ and\
  \bibinfo {author} {\bibfnamefont {M.}~\bibnamefont {Shafer}},\ }\href
  {\doibase 10.1103/PhysRevLett.37.1407} {\bibfield  {journal} {\bibinfo
  {journal} {Phys. Rev. Lett.}\ }\textbf {\bibinfo {volume} {37}},\ \bibinfo
  {pages} {1407} (\bibinfo {year} {1976})}\BibitemShut {NoStop}%
\bibitem [{\citenamefont {Sugai}\ \emph {et~al.}(2006)\citenamefont {Sugai},
  \citenamefont {Takayanagi},\ and\ \citenamefont
  {Hayamizu}}]{sugai_phason_2006}%
  \BibitemOpen
  \bibfield  {author} {\bibinfo {author} {\bibfnamefont {S.}~\bibnamefont
  {Sugai}}, \bibinfo {author} {\bibfnamefont {Y.}~\bibnamefont {Takayanagi}}, \
  and\ \bibinfo {author} {\bibfnamefont {N.}~\bibnamefont {Hayamizu}},\ }\href
  {\doibase 10.1103/PhysRevLett.96.137003} {\bibfield  {journal} {\bibinfo
  {journal} {Phys. Rev. Lett.}\ }\textbf {\bibinfo {volume} {96}},\ \bibinfo
  {pages} {137003} (\bibinfo {year} {2006})}\BibitemShut {NoStop}%
\bibitem [{\citenamefont {Travaglini}\ \emph {et~al.}(1983)\citenamefont
  {Travaglini}, \citenamefont {Morke},\ and\ \citenamefont
  {Wachter}}]{travaglini_CDW_1983}%
  \BibitemOpen
  \bibfield  {author} {\bibinfo {author} {\bibfnamefont {G.}~\bibnamefont
  {Travaglini}}, \bibinfo {author} {\bibfnamefont {I.}~\bibnamefont {Morke}}, \
  and\ \bibinfo {author} {\bibfnamefont {P.}~\bibnamefont {Wachter}},\ }\href
  {\doibase https://doi.org/10.1016/0038-1098(83)90483-0} {\bibfield  {journal}
  {\bibinfo  {journal} {Solid State Commun.}\ }\textbf {\bibinfo {volume}
  {45}},\ \bibinfo {pages} {289 } (\bibinfo {year} {1983})}\BibitemShut
  {NoStop}%
\bibitem [{\citenamefont {Cea}\ \emph {et~al.}(2016)\citenamefont {Cea},
  \citenamefont {Castellani},\ and\ \citenamefont {Benfatto}}]{cea_thg_prb16}%
  \BibitemOpen
  \bibfield  {author} {\bibinfo {author} {\bibfnamefont {T.}~\bibnamefont
  {Cea}}, \bibinfo {author} {\bibfnamefont {C.}~\bibnamefont {Castellani}}, \
  and\ \bibinfo {author} {\bibfnamefont {L.}~\bibnamefont {Benfatto}},\ }\href
  {\doibase 10.1103/PhysRevB.93.180507} {\bibfield  {journal} {\bibinfo
  {journal} {Phys. Rev. B}\ }\textbf {\bibinfo {volume} {93}},\ \bibinfo
  {pages} {180507} (\bibinfo {year} {2016})}\BibitemShut {NoStop}%
\bibitem [{\citenamefont {Volkov}\ and\ \citenamefont
  {Kogan}(1974)}]{volkov73}%
  \BibitemOpen
  \bibfield  {author} {\bibinfo {author} {\bibfnamefont {A.~F.}\ \bibnamefont
  {Volkov}}\ and\ \bibinfo {author} {\bibfnamefont {S.~M.}\ \bibnamefont
  {Kogan}},\ }\href@noop {} {\bibfield  {journal} {\bibinfo  {journal} {Zh.
  Eksp. Teor. Fiz. [Sov. Phys. JETP]}\ }\textbf {\bibinfo {volume} {65[38]}},\
  \bibinfo {pages} {2038[1018]} (\bibinfo {year} {1973[1974]})}\BibitemShut
  {NoStop}%
\bibitem [{\citenamefont {Kulik}\ \emph {et~al.}(1981)\citenamefont {Kulik},
  \citenamefont {Entin-Wohlman},\ and\ \citenamefont {Orbach}}]{kulik81}%
  \BibitemOpen
  \bibfield  {author} {\bibinfo {author} {\bibfnamefont {I.~O.}\ \bibnamefont
  {Kulik}}, \bibinfo {author} {\bibfnamefont {O.}~\bibnamefont
  {Entin-Wohlman}}, \ and\ \bibinfo {author} {\bibfnamefont {R.}~\bibnamefont
  {Orbach}},\ }\href {\doibase 10.1007/BF00115617} {\bibfield  {journal}
  {\bibinfo  {journal} {J. Low Temp. Phys.}\ }\textbf {\bibinfo {volume}
  {43}},\ \bibinfo {pages} {591} (\bibinfo {year} {1981})}\BibitemShut
  {NoStop}%
\bibitem [{\citenamefont {Cea}\ and\ \citenamefont
  {Benfatto}(2014)}]{cea_nature_2014}%
  \BibitemOpen
  \bibfield  {author} {\bibinfo {author} {\bibfnamefont {T.}~\bibnamefont
  {Cea}}\ and\ \bibinfo {author} {\bibfnamefont {L.}~\bibnamefont {Benfatto}},\
  }\href {\doibase 10.1103/PhysRevB.90.224515} {\bibfield  {journal} {\bibinfo
  {journal} {Phys. Rev. B}\ }\textbf {\bibinfo {volume} {90}},\ \bibinfo
  {pages} {224515} (\bibinfo {year} {2014})}\BibitemShut {NoStop}%
\bibitem [{\citenamefont {Cea}\ \emph {et~al.}(2015)\citenamefont {Cea},
  \citenamefont {Castellani}, \citenamefont {Seibold},\ and\ \citenamefont
  {Benfatto}}]{cea_prl15}%
  \BibitemOpen
  \bibfield  {author} {\bibinfo {author} {\bibfnamefont {T.}~\bibnamefont
  {Cea}}, \bibinfo {author} {\bibfnamefont {C.}~\bibnamefont {Castellani}},
  \bibinfo {author} {\bibfnamefont {G.}~\bibnamefont {Seibold}}, \ and\
  \bibinfo {author} {\bibfnamefont {L.}~\bibnamefont {Benfatto}},\ }\href
  {\doibase 10.1103/PhysRevLett.115.157002} {\bibfield  {journal} {\bibinfo
  {journal} {Phys. Rev. Lett.}\ }\textbf {\bibinfo {volume} {115}},\ \bibinfo
  {pages} {157002} (\bibinfo {year} {2015})}\BibitemShut {NoStop}%
\bibitem [{\citenamefont {Matsunaga}\ \emph {et~al.}(2013)\citenamefont
  {Matsunaga}, \citenamefont {Hamada}, \citenamefont {Makise}, \citenamefont
  {Uzawa}, \citenamefont {Terai}, \citenamefont {Wang},\ and\ \citenamefont
  {Shimano}}]{shimano_prl13}%
  \BibitemOpen
  \bibfield  {author} {\bibinfo {author} {\bibfnamefont {R.}~\bibnamefont
  {Matsunaga}}, \bibinfo {author} {\bibfnamefont {Y.~I.}\ \bibnamefont
  {Hamada}}, \bibinfo {author} {\bibfnamefont {K.}~\bibnamefont {Makise}},
  \bibinfo {author} {\bibfnamefont {Y.}~\bibnamefont {Uzawa}}, \bibinfo
  {author} {\bibfnamefont {H.}~\bibnamefont {Terai}}, \bibinfo {author}
  {\bibfnamefont {Z.}~\bibnamefont {Wang}}, \ and\ \bibinfo {author}
  {\bibfnamefont {R.}~\bibnamefont {Shimano}},\ }\href {\doibase
  10.1103/PhysRevLett.111.057002} {\bibfield  {journal} {\bibinfo  {journal}
  {Phys. Rev. Lett.}\ }\textbf {\bibinfo {volume} {111}},\ \bibinfo {pages}
  {057002} (\bibinfo {year} {2013})}\BibitemShut {NoStop}%
\bibitem [{\citenamefont {Matsunaga}\ \emph {et~al.}(2014)\citenamefont
  {Matsunaga}, \citenamefont {Tsuji}, \citenamefont {Fujita}, \citenamefont
  {Sugioka}, \citenamefont {Makise}, \citenamefont {Uzawa}, \citenamefont
  {Terai}, \citenamefont {Wang}, \citenamefont {Aoki},\ and\ \citenamefont
  {Shimano}}]{shimano_science14}%
  \BibitemOpen
  \bibfield  {author} {\bibinfo {author} {\bibfnamefont {R.}~\bibnamefont
  {Matsunaga}}, \bibinfo {author} {\bibfnamefont {N.}~\bibnamefont {Tsuji}},
  \bibinfo {author} {\bibfnamefont {H.}~\bibnamefont {Fujita}}, \bibinfo
  {author} {\bibfnamefont {A.}~\bibnamefont {Sugioka}}, \bibinfo {author}
  {\bibfnamefont {K.}~\bibnamefont {Makise}}, \bibinfo {author} {\bibfnamefont
  {Y.}~\bibnamefont {Uzawa}}, \bibinfo {author} {\bibfnamefont
  {H.}~\bibnamefont {Terai}}, \bibinfo {author} {\bibfnamefont
  {Z.}~\bibnamefont {Wang}}, \bibinfo {author} {\bibfnamefont {H.}~\bibnamefont
  {Aoki}}, \ and\ \bibinfo {author} {\bibfnamefont {R.}~\bibnamefont
  {Shimano}},\ }\href {\doibase 10.1126/science.1254697} {\bibfield  {journal}
  {\bibinfo  {journal} {Science}\ }\textbf {\bibinfo {volume} {345}},\ \bibinfo
  {pages} {1145} (\bibinfo {year} {2014})}\BibitemShut {NoStop}%
\bibitem [{\citenamefont {Katsumi}\ \emph {et~al.}(2018)\citenamefont
  {Katsumi}, \citenamefont {Tsuji}, \citenamefont {Hamada}, \citenamefont
  {Matsunaga}, \citenamefont {Schneeloch}, \citenamefont {Zhong}, \citenamefont
  {Gu}, \citenamefont {Aoki}, \citenamefont {Gallais},\ and\ \citenamefont
  {Shimano}}]{PhysRevLett.120.117001}%
  \BibitemOpen
  \bibfield  {author} {\bibinfo {author} {\bibfnamefont {K.}~\bibnamefont
  {Katsumi}}, \bibinfo {author} {\bibfnamefont {N.}~\bibnamefont {Tsuji}},
  \bibinfo {author} {\bibfnamefont {Y.~I.}\ \bibnamefont {Hamada}}, \bibinfo
  {author} {\bibfnamefont {R.}~\bibnamefont {Matsunaga}}, \bibinfo {author}
  {\bibfnamefont {J.}~\bibnamefont {Schneeloch}}, \bibinfo {author}
  {\bibfnamefont {R.~D.}\ \bibnamefont {Zhong}}, \bibinfo {author}
  {\bibfnamefont {G.~D.}\ \bibnamefont {Gu}}, \bibinfo {author} {\bibfnamefont
  {H.}~\bibnamefont {Aoki}}, \bibinfo {author} {\bibfnamefont {Y.}~\bibnamefont
  {Gallais}}, \ and\ \bibinfo {author} {\bibfnamefont {R.}~\bibnamefont
  {Shimano}},\ }\href {\doibase 10.1103/PhysRevLett.120.117001} {\bibfield
  {journal} {\bibinfo  {journal} {Phys. Rev. Lett.}\ }\textbf {\bibinfo
  {volume} {120}},\ \bibinfo {pages} {117001} (\bibinfo {year}
  {2018})}\BibitemShut {NoStop}%
\bibitem [{\citenamefont {Sherman}\ \emph {et~al.}(2015)\citenamefont
  {Sherman}, \citenamefont {Pracht}, \citenamefont {Gorshunov}, \citenamefont
  {Poran}, \citenamefont {Jesudasan}, \citenamefont {Chand}, \citenamefont
  {Raychaudhuri}, \citenamefont {Swanson}, \citenamefont {Trivedi},
  \citenamefont {Auerbach}, \citenamefont {Scheffler}, \citenamefont
  {Frydman},\ and\ \citenamefont {Dressel}}]{frydman_natphys15}%
  \BibitemOpen
  \bibfield  {author} {\bibinfo {author} {\bibfnamefont {D.}~\bibnamefont
  {Sherman}}, \bibinfo {author} {\bibfnamefont {U.~S.}\ \bibnamefont {Pracht}},
  \bibinfo {author} {\bibfnamefont {B.}~\bibnamefont {Gorshunov}}, \bibinfo
  {author} {\bibfnamefont {S.}~\bibnamefont {Poran}}, \bibinfo {author}
  {\bibfnamefont {J.}~\bibnamefont {Jesudasan}}, \bibinfo {author}
  {\bibfnamefont {M.}~\bibnamefont {Chand}}, \bibinfo {author} {\bibfnamefont
  {P.}~\bibnamefont {Raychaudhuri}}, \bibinfo {author} {\bibfnamefont
  {M.}~\bibnamefont {Swanson}}, \bibinfo {author} {\bibfnamefont
  {N.}~\bibnamefont {Trivedi}}, \bibinfo {author} {\bibfnamefont
  {A.}~\bibnamefont {Auerbach}}, \bibinfo {author} {\bibfnamefont
  {M.}~\bibnamefont {Scheffler}}, \bibinfo {author} {\bibfnamefont
  {A.}~\bibnamefont {Frydman}}, \ and\ \bibinfo {author} {\bibfnamefont
  {M.}~\bibnamefont {Dressel}},\ }\href {http://dx.doi.org/10.1038/nphys3227}
  {\bibfield  {journal} {\bibinfo  {journal} {Nat. Phys.}\ }\textbf {\bibinfo
  {volume} {11}},\ \bibinfo {pages} {188} (\bibinfo {year} {2015})}\BibitemShut
  {NoStop}%
\bibitem [{\citenamefont {Pekker}\ and\ \citenamefont
  {Varma}(2015)}]{pekker_amplitudehiggs_2015}%
  \BibitemOpen
  \bibfield  {author} {\bibinfo {author} {\bibfnamefont {D.}~\bibnamefont
  {Pekker}}\ and\ \bibinfo {author} {\bibfnamefont {C.}~\bibnamefont {Varma}},\
  }\href {\doibase 10.1146/annurev-conmatphys-031214-014350} {\bibfield
  {journal} {\bibinfo  {journal} {Annu. Rev. Condens. Matter Phys.}\ }\textbf
  {\bibinfo {volume} {6}},\ \bibinfo {pages} {269} (\bibinfo {year}
  {2015})}\BibitemShut {NoStop}%
\bibitem [{\citenamefont {Cheng}\ \emph {et~al.}(2016)\citenamefont {Cheng},
  \citenamefont {Wu}, \citenamefont {Laurita}, \citenamefont {Singh},
  \citenamefont {Chand}, \citenamefont {Raychaudhuri},\ and\ \citenamefont
  {Armitage}}]{armitage_prb16}%
  \BibitemOpen
  \bibfield  {author} {\bibinfo {author} {\bibfnamefont {B.}~\bibnamefont
  {Cheng}}, \bibinfo {author} {\bibfnamefont {L.}~\bibnamefont {Wu}}, \bibinfo
  {author} {\bibfnamefont {N.~J.}\ \bibnamefont {Laurita}}, \bibinfo {author}
  {\bibfnamefont {H.}~\bibnamefont {Singh}}, \bibinfo {author} {\bibfnamefont
  {M.}~\bibnamefont {Chand}}, \bibinfo {author} {\bibfnamefont
  {P.}~\bibnamefont {Raychaudhuri}}, \ and\ \bibinfo {author} {\bibfnamefont
  {N.~P.}\ \bibnamefont {Armitage}},\ }\href {\doibase
  10.1103/PhysRevB.93.180511} {\bibfield  {journal} {\bibinfo  {journal} {Phys.
  Rev. B}\ }\textbf {\bibinfo {volume} {93}},\ \bibinfo {pages} {180511}
  (\bibinfo {year} {2016})}\BibitemShut {NoStop}%
\bibitem [{\citenamefont {Tsuji}\ \emph {et~al.}(2016)\citenamefont {Tsuji},
  \citenamefont {Murakami},\ and\ \citenamefont {Aoki}}]{tsuji_prb16}%
  \BibitemOpen
  \bibfield  {author} {\bibinfo {author} {\bibfnamefont {N.}~\bibnamefont
  {Tsuji}}, \bibinfo {author} {\bibfnamefont {Y.}~\bibnamefont {Murakami}}, \
  and\ \bibinfo {author} {\bibfnamefont {H.}~\bibnamefont {Aoki}},\ }\href
  {\doibase 10.1103/PhysRevB.94.224519} {\bibfield  {journal} {\bibinfo
  {journal} {Phys. Rev. B}\ }\textbf {\bibinfo {volume} {94}},\ \bibinfo
  {pages} {224519} (\bibinfo {year} {2016})}\BibitemShut {NoStop}%
\bibitem [{\citenamefont {Matsunaga}\ \emph {et~al.}(2017)\citenamefont
  {Matsunaga}, \citenamefont {Tsuji}, \citenamefont {Makise}, \citenamefont
  {Terai}, \citenamefont {Aoki},\ and\ \citenamefont
  {Shimano}}]{PhysRevB.96.020505}%
  \BibitemOpen
  \bibfield  {author} {\bibinfo {author} {\bibfnamefont {R.}~\bibnamefont
  {Matsunaga}}, \bibinfo {author} {\bibfnamefont {N.}~\bibnamefont {Tsuji}},
  \bibinfo {author} {\bibfnamefont {K.}~\bibnamefont {Makise}}, \bibinfo
  {author} {\bibfnamefont {H.}~\bibnamefont {Terai}}, \bibinfo {author}
  {\bibfnamefont {H.}~\bibnamefont {Aoki}}, \ and\ \bibinfo {author}
  {\bibfnamefont {R.}~\bibnamefont {Shimano}},\ }\href {\doibase
  10.1103/PhysRevB.96.020505} {\bibfield  {journal} {\bibinfo  {journal} {Phys.
  Rev. B}\ }\textbf {\bibinfo {volume} {96}},\ \bibinfo {pages} {020505}
  (\bibinfo {year} {2017})}\BibitemShut {NoStop}%
\bibitem [{\citenamefont {Cea}\ \emph {et~al.}(2018)\citenamefont {Cea},
  \citenamefont {Barone}, \citenamefont {Castellani},\ and\ \citenamefont
  {Benfatto}}]{PhysRevB.97.094516}%
  \BibitemOpen
  \bibfield  {author} {\bibinfo {author} {\bibfnamefont {T.}~\bibnamefont
  {Cea}}, \bibinfo {author} {\bibfnamefont {P.}~\bibnamefont {Barone}},
  \bibinfo {author} {\bibfnamefont {C.}~\bibnamefont {Castellani}}, \ and\
  \bibinfo {author} {\bibfnamefont {L.}~\bibnamefont {Benfatto}},\ }\href
  {\doibase 10.1103/PhysRevB.97.094516} {\bibfield  {journal} {\bibinfo
  {journal} {Phys. Rev. B}\ }\textbf {\bibinfo {volume} {97}},\ \bibinfo
  {pages} {094516} (\bibinfo {year} {2018})}\BibitemShut {NoStop}%
\bibitem [{\citenamefont {Sooryakumar}\ and\ \citenamefont
  {Klein}(1980)}]{sooryakumar_raman_1980}%
  \BibitemOpen
  \bibfield  {author} {\bibinfo {author} {\bibfnamefont {R.}~\bibnamefont
  {Sooryakumar}}\ and\ \bibinfo {author} {\bibfnamefont {M.}~\bibnamefont
  {Klein}},\ }\href {\doibase 10.1103/PhysRevLett.45.660} {\bibfield  {journal}
  {\bibinfo  {journal} {Phys. Rev. Lett.}\ }\textbf {\bibinfo {volume} {45}},\
  \bibinfo {pages} {660} (\bibinfo {year} {1980})}\BibitemShut {NoStop}%
\bibitem [{\citenamefont {M\'easson}\ \emph {et~al.}(2014)\citenamefont
  {M\'easson}, \citenamefont {Gallais}, \citenamefont {Cazayous}, \citenamefont
  {Clair}, \citenamefont {Rodi\`ere}, \citenamefont {Cario},\ and\
  \citenamefont {Sacuto}}]{measson_amplitude_2014}%
  \BibitemOpen
  \bibfield  {author} {\bibinfo {author} {\bibfnamefont {M.-A.}\ \bibnamefont
  {M\'easson}}, \bibinfo {author} {\bibfnamefont {Y.}~\bibnamefont {Gallais}},
  \bibinfo {author} {\bibfnamefont {M.}~\bibnamefont {Cazayous}}, \bibinfo
  {author} {\bibfnamefont {B.}~\bibnamefont {Clair}}, \bibinfo {author}
  {\bibfnamefont {P.}~\bibnamefont {Rodi\`ere}}, \bibinfo {author}
  {\bibfnamefont {L.}~\bibnamefont {Cario}}, \ and\ \bibinfo {author}
  {\bibfnamefont {A.}~\bibnamefont {Sacuto}},\ }\href {\doibase
  10.1103/PhysRevB.89.060503} {\bibfield  {journal} {\bibinfo  {journal} {Phys.
  Rev. B}\ }\textbf {\bibinfo {volume} {89}},\ \bibinfo {pages} {060503}
  (\bibinfo {year} {2014})}\BibitemShut {NoStop}%
\bibitem [{\citenamefont {Grasset}\ \emph {et~al.}(2018)\citenamefont
  {Grasset}, \citenamefont {Cea}, \citenamefont {Gallais}, \citenamefont
  {Cazayous}, \citenamefont {Sacuto}, \citenamefont {Cario}, \citenamefont
  {Benfatto},\ and\ \citenamefont {M\'easson}}]{grasset_prb_2018}%
  \BibitemOpen
  \bibfield  {author} {\bibinfo {author} {\bibfnamefont {R.}~\bibnamefont
  {Grasset}}, \bibinfo {author} {\bibfnamefont {T.}~\bibnamefont {Cea}},
  \bibinfo {author} {\bibfnamefont {Y.}~\bibnamefont {Gallais}}, \bibinfo
  {author} {\bibfnamefont {M.}~\bibnamefont {Cazayous}}, \bibinfo {author}
  {\bibfnamefont {A.}~\bibnamefont {Sacuto}}, \bibinfo {author} {\bibfnamefont
  {L.}~\bibnamefont {Cario}}, \bibinfo {author} {\bibfnamefont
  {L.}~\bibnamefont {Benfatto}}, \ and\ \bibinfo {author} {\bibfnamefont
  {M.-A.}\ \bibnamefont {M\'easson}},\ }\href {\doibase
  10.1103/PhysRevB.97.094502} {\bibfield  {journal} {\bibinfo  {journal} {Phys.
  Rev. B}\ }\textbf {\bibinfo {volume} {97}},\ \bibinfo {pages} {094502}
  (\bibinfo {year} {2018})}\BibitemShut {NoStop}%
\bibitem [{\citenamefont {Browne}\ and\ \citenamefont
  {Levin}(1983)}]{browne_prb83}%
  \BibitemOpen
  \bibfield  {author} {\bibinfo {author} {\bibfnamefont {D.~A.}\ \bibnamefont
  {Browne}}\ and\ \bibinfo {author} {\bibfnamefont {K.}~\bibnamefont {Levin}},\
  }\href {\doibase 10.1103/PhysRevB.28.4029} {\bibfield  {journal} {\bibinfo
  {journal} {Phys. Rev. B}\ }\textbf {\bibinfo {volume} {28}},\ \bibinfo
  {pages} {4029} (\bibinfo {year} {1983})}\BibitemShut {NoStop}%
\bibitem [{\citenamefont {Ghiringhelli}\ \emph {et~al.}(2012)\citenamefont
  {Ghiringhelli}, \citenamefont {Le~Tacon}, \citenamefont {Minola},
  \citenamefont {Blanco-Canosa}, \citenamefont {Mazzoli}, \citenamefont
  {Brookes}, \citenamefont {De~Luca}, \citenamefont {Frano}, \citenamefont
  {Hawthorn}, \citenamefont {He}, \citenamefont {Loew}, \citenamefont {Sala},
  \citenamefont {Peets}, \citenamefont {Salluzzo}, \citenamefont {Schierle},
  \citenamefont {Sutarto}, \citenamefont {Sawatzky}, \citenamefont {Weschke},
  \citenamefont {Keimer},\ and\ \citenamefont
  {Braicovich}}]{ghiringhelli_long-range_2012}%
  \BibitemOpen
  \bibfield  {author} {\bibinfo {author} {\bibfnamefont {G.}~\bibnamefont
  {Ghiringhelli}}, \bibinfo {author} {\bibfnamefont {M.}~\bibnamefont
  {Le~Tacon}}, \bibinfo {author} {\bibfnamefont {M.}~\bibnamefont {Minola}},
  \bibinfo {author} {\bibfnamefont {S.}~\bibnamefont {Blanco-Canosa}}, \bibinfo
  {author} {\bibfnamefont {C.}~\bibnamefont {Mazzoli}}, \bibinfo {author}
  {\bibfnamefont {N.~B.}\ \bibnamefont {Brookes}}, \bibinfo {author}
  {\bibfnamefont {G.~M.}\ \bibnamefont {De~Luca}}, \bibinfo {author}
  {\bibfnamefont {A.}~\bibnamefont {Frano}}, \bibinfo {author} {\bibfnamefont
  {D.~G.}\ \bibnamefont {Hawthorn}}, \bibinfo {author} {\bibfnamefont
  {F.}~\bibnamefont {He}}, \bibinfo {author} {\bibfnamefont {T.}~\bibnamefont
  {Loew}}, \bibinfo {author} {\bibfnamefont {M.~M.}\ \bibnamefont {Sala}},
  \bibinfo {author} {\bibfnamefont {D.~C.}\ \bibnamefont {Peets}}, \bibinfo
  {author} {\bibfnamefont {M.}~\bibnamefont {Salluzzo}}, \bibinfo {author}
  {\bibfnamefont {E.}~\bibnamefont {Schierle}}, \bibinfo {author}
  {\bibfnamefont {R.}~\bibnamefont {Sutarto}}, \bibinfo {author} {\bibfnamefont
  {G.~A.}\ \bibnamefont {Sawatzky}}, \bibinfo {author} {\bibfnamefont
  {E.}~\bibnamefont {Weschke}}, \bibinfo {author} {\bibfnamefont
  {B.}~\bibnamefont {Keimer}}, \ and\ \bibinfo {author} {\bibfnamefont
  {L.}~\bibnamefont {Braicovich}},\ }\href {\doibase 10.1126/science.1223532}
  {\bibfield  {journal} {\bibinfo  {journal} {Science}\ }\textbf {\bibinfo
  {volume} {337}},\ \bibinfo {pages} {821} (\bibinfo {year}
  {2012})}\BibitemShut {NoStop}%
\bibitem [{\citenamefont {da~Silva~Neto}\ \emph {et~al.}(2014)\citenamefont
  {da~Silva~Neto}, \citenamefont {Aynajian}, \citenamefont {Frano},
  \citenamefont {Comin}, \citenamefont {Schierle}, \citenamefont {Weschke},
  \citenamefont {Gyenis}, \citenamefont {Wen}, \citenamefont {Schneeloch},
  \citenamefont {Xu}, \citenamefont {Ono}, \citenamefont {Gu}, \citenamefont
  {Le~Tacon},\ and\ \citenamefont {Yazdani}}]{da_silva_neto_ubiquitous_2014}%
  \BibitemOpen
  \bibfield  {author} {\bibinfo {author} {\bibfnamefont {E.~H.}\ \bibnamefont
  {da~Silva~Neto}}, \bibinfo {author} {\bibfnamefont {P.}~\bibnamefont
  {Aynajian}}, \bibinfo {author} {\bibfnamefont {A.}~\bibnamefont {Frano}},
  \bibinfo {author} {\bibfnamefont {R.}~\bibnamefont {Comin}}, \bibinfo
  {author} {\bibfnamefont {E.}~\bibnamefont {Schierle}}, \bibinfo {author}
  {\bibfnamefont {E.}~\bibnamefont {Weschke}}, \bibinfo {author} {\bibfnamefont
  {A.}~\bibnamefont {Gyenis}}, \bibinfo {author} {\bibfnamefont
  {J.}~\bibnamefont {Wen}}, \bibinfo {author} {\bibfnamefont {J.}~\bibnamefont
  {Schneeloch}}, \bibinfo {author} {\bibfnamefont {Z.}~\bibnamefont {Xu}},
  \bibinfo {author} {\bibfnamefont {S.}~\bibnamefont {Ono}}, \bibinfo {author}
  {\bibfnamefont {G.}~\bibnamefont {Gu}}, \bibinfo {author} {\bibfnamefont
  {M.}~\bibnamefont {Le~Tacon}}, \ and\ \bibinfo {author} {\bibfnamefont
  {A.}~\bibnamefont {Yazdani}},\ }\href {\doibase 10.1126/science.1243479}
  {\bibfield  {journal} {\bibinfo  {journal} {Science}\ }\textbf {\bibinfo
  {volume} {343}},\ \bibinfo {pages} {393} (\bibinfo {year}
  {2014})}\BibitemShut {NoStop}%
\bibitem [{\citenamefont {Comin}\ \emph {et~al.}(2014)\citenamefont {Comin},
  \citenamefont {Frano}, \citenamefont {Yee}, \citenamefont {Yoshida},
  \citenamefont {Eisaki}, \citenamefont {Schierle}, \citenamefont {Weschke},
  \citenamefont {Sutarto}, \citenamefont {He}, \citenamefont {Soumyanarayanan},
  \citenamefont {He}, \citenamefont {Le~Tacon}, \citenamefont {Elfimov},
  \citenamefont {Hoffman}, \citenamefont {Sawatzky}, \citenamefont {Keimer},\
  and\ \citenamefont {Damascelli}}]{comin_charge_2014}%
  \BibitemOpen
  \bibfield  {author} {\bibinfo {author} {\bibfnamefont {R.}~\bibnamefont
  {Comin}}, \bibinfo {author} {\bibfnamefont {A.}~\bibnamefont {Frano}},
  \bibinfo {author} {\bibfnamefont {M.~M.}\ \bibnamefont {Yee}}, \bibinfo
  {author} {\bibfnamefont {Y.}~\bibnamefont {Yoshida}}, \bibinfo {author}
  {\bibfnamefont {H.}~\bibnamefont {Eisaki}}, \bibinfo {author} {\bibfnamefont
  {E.}~\bibnamefont {Schierle}}, \bibinfo {author} {\bibfnamefont
  {E.}~\bibnamefont {Weschke}}, \bibinfo {author} {\bibfnamefont
  {R.}~\bibnamefont {Sutarto}}, \bibinfo {author} {\bibfnamefont
  {F.}~\bibnamefont {He}}, \bibinfo {author} {\bibfnamefont {A.}~\bibnamefont
  {Soumyanarayanan}}, \bibinfo {author} {\bibfnamefont {Y.}~\bibnamefont {He}},
  \bibinfo {author} {\bibfnamefont {M.}~\bibnamefont {Le~Tacon}}, \bibinfo
  {author} {\bibfnamefont {I.~S.}\ \bibnamefont {Elfimov}}, \bibinfo {author}
  {\bibfnamefont {J.~E.}\ \bibnamefont {Hoffman}}, \bibinfo {author}
  {\bibfnamefont {G.~A.}\ \bibnamefont {Sawatzky}}, \bibinfo {author}
  {\bibfnamefont {B.}~\bibnamefont {Keimer}}, \ and\ \bibinfo {author}
  {\bibfnamefont {A.}~\bibnamefont {Damascelli}},\ }\href {\doibase
  10.1126/science.1242996} {\bibfield  {journal} {\bibinfo  {journal}
  {Science}\ }\textbf {\bibinfo {volume} {343}},\ \bibinfo {pages} {390}
  (\bibinfo {year} {2014})}\BibitemShut {NoStop}%
\bibitem [{\citenamefont {Wu}\ \emph {et~al.}(2011)\citenamefont {Wu},
  \citenamefont {Mayaffre}, \citenamefont {Kr\"amer}, \citenamefont
  {Horvati\'c}, \citenamefont {Berthier}, \citenamefont {Hardy}, \citenamefont
  {Liang}, \citenamefont {Bonn},\ and\ \citenamefont
  {Julien}}]{wu_magnetic-field-induced_2011}%
  \BibitemOpen
  \bibfield  {author} {\bibinfo {author} {\bibfnamefont {T.}~\bibnamefont
  {Wu}}, \bibinfo {author} {\bibfnamefont {H.}~\bibnamefont {Mayaffre}},
  \bibinfo {author} {\bibfnamefont {S.}~\bibnamefont {Kr\"amer}}, \bibinfo
  {author} {\bibfnamefont {M.}~\bibnamefont {Horvati\'c}}, \bibinfo {author}
  {\bibfnamefont {C.}~\bibnamefont {Berthier}}, \bibinfo {author}
  {\bibfnamefont {W.~N.}\ \bibnamefont {Hardy}}, \bibinfo {author}
  {\bibfnamefont {R.}~\bibnamefont {Liang}}, \bibinfo {author} {\bibfnamefont
  {D.~A.}\ \bibnamefont {Bonn}}, \ and\ \bibinfo {author} {\bibfnamefont
  {M.-H.}\ \bibnamefont {Julien}},\ }\href {\doibase 10.1038/nature10345}
  {\bibfield  {journal} {\bibinfo  {journal} {Nature}\ }\textbf {\bibinfo
  {volume} {477}},\ \bibinfo {pages} {191} (\bibinfo {year}
  {2011})}\BibitemShut {NoStop}%
\bibitem [{\citenamefont {Tranquada}\ \emph {et~al.}(1995)\citenamefont
  {Tranquada}, \citenamefont {Sternlieb}, \citenamefont {Axe}, \citenamefont
  {Nakamura},\ and\ \citenamefont {Uchida}}]{tranquada_evidence_1995}%
  \BibitemOpen
  \bibfield  {author} {\bibinfo {author} {\bibfnamefont {J.~M.}\ \bibnamefont
  {Tranquada}}, \bibinfo {author} {\bibfnamefont {B.~J.}\ \bibnamefont
  {Sternlieb}}, \bibinfo {author} {\bibfnamefont {J.~D.}\ \bibnamefont {Axe}},
  \bibinfo {author} {\bibfnamefont {Y.}~\bibnamefont {Nakamura}}, \ and\
  \bibinfo {author} {\bibfnamefont {S.}~\bibnamefont {Uchida}},\ }\href
  {\doibase 10.1038/375561a0} {\bibfield  {journal} {\bibinfo  {journal}
  {Nature}\ }\textbf {\bibinfo {volume} {375}},\ \bibinfo {pages} {561}
  (\bibinfo {year} {1995})}\BibitemShut {NoStop}%
\bibitem [{\citenamefont {Lake}\ \emph {et~al.}(2002)\citenamefont {Lake},
  \citenamefont {Ronnow}, \citenamefont {Christensen}, \citenamefont {Aeppli},
  \citenamefont {Lefmann}, \citenamefont {McMorrow}, \citenamefont
  {Vorderwisch}, \citenamefont {Smeibidl}, \citenamefont {Mangkorntong},
  \citenamefont {Sasagawa}, \citenamefont {Nohara}, \citenamefont {Takagi},\
  and\ \citenamefont {Mason}}]{lake_antiferromagnetic_2002}%
  \BibitemOpen
  \bibfield  {author} {\bibinfo {author} {\bibfnamefont {B.}~\bibnamefont
  {Lake}}, \bibinfo {author} {\bibfnamefont {H.~M.}\ \bibnamefont {Ronnow}},
  \bibinfo {author} {\bibfnamefont {N.~B.}\ \bibnamefont {Christensen}},
  \bibinfo {author} {\bibfnamefont {G.}~\bibnamefont {Aeppli}}, \bibinfo
  {author} {\bibfnamefont {K.}~\bibnamefont {Lefmann}}, \bibinfo {author}
  {\bibfnamefont {D.~F.}\ \bibnamefont {McMorrow}}, \bibinfo {author}
  {\bibfnamefont {P.}~\bibnamefont {Vorderwisch}}, \bibinfo {author}
  {\bibfnamefont {P.}~\bibnamefont {Smeibidl}}, \bibinfo {author}
  {\bibfnamefont {N.}~\bibnamefont {Mangkorntong}}, \bibinfo {author}
  {\bibfnamefont {T.}~\bibnamefont {Sasagawa}}, \bibinfo {author}
  {\bibfnamefont {M.}~\bibnamefont {Nohara}}, \bibinfo {author} {\bibfnamefont
  {H.}~\bibnamefont {Takagi}}, \ and\ \bibinfo {author} {\bibfnamefont {T.~E.}\
  \bibnamefont {Mason}},\ }\href {\doibase 10.1038/415299a} {\bibfield
  {journal} {\bibinfo  {journal} {Nature}\ }\textbf {\bibinfo {volume} {415}},\
  \bibinfo {pages} {299} (\bibinfo {year} {2002})}\BibitemShut {NoStop}%
\bibitem [{\citenamefont {Sipos}\ \emph {et~al.}(2008)\citenamefont {Sipos},
  \citenamefont {Kusmartseva}, \citenamefont {Akrap}, \citenamefont {Berger},
  \citenamefont {Forr\'o},\ and\ \citenamefont {Tuti\v{s}}}]{sipos_mott_2008}%
  \BibitemOpen
  \bibfield  {author} {\bibinfo {author} {\bibfnamefont {B.}~\bibnamefont
  {Sipos}}, \bibinfo {author} {\bibfnamefont {A.~F.}\ \bibnamefont
  {Kusmartseva}}, \bibinfo {author} {\bibfnamefont {A.}~\bibnamefont {Akrap}},
  \bibinfo {author} {\bibfnamefont {H.}~\bibnamefont {Berger}}, \bibinfo
  {author} {\bibfnamefont {L.}~\bibnamefont {Forr\'o}}, \ and\ \bibinfo
  {author} {\bibfnamefont {E.}~\bibnamefont {Tuti\v{s}}},\ }\href {\doibase
  10.1038/nmat2318} {\bibfield  {journal} {\bibinfo  {journal} {Nature Mater.}\
  }\textbf {\bibinfo {volume} {7}},\ \bibinfo {pages} {960} (\bibinfo {year}
  {2008})}\BibitemShut {NoStop}%
\bibitem [{\citenamefont {Kusmartseva}\ \emph {et~al.}(2009)\citenamefont
  {Kusmartseva}, \citenamefont {Sipos}, \citenamefont {Berger}, \citenamefont
  {Forr\'o},\ and\ \citenamefont {Tuti\ifmmode~\check{s}\else
  \v{s}\fi{}}}]{kusmartseva_pressure_2009}%
  \BibitemOpen
  \bibfield  {author} {\bibinfo {author} {\bibfnamefont {A.~F.}\ \bibnamefont
  {Kusmartseva}}, \bibinfo {author} {\bibfnamefont {B.}~\bibnamefont {Sipos}},
  \bibinfo {author} {\bibfnamefont {H.}~\bibnamefont {Berger}}, \bibinfo
  {author} {\bibfnamefont {L.}~\bibnamefont {Forr\'o}}, \ and\ \bibinfo
  {author} {\bibfnamefont {E.}~\bibnamefont {Tuti\ifmmode~\check{s}\else
  \v{s}\fi{}}},\ }\href {\doibase 10.1103/PhysRevLett.103.236401} {\bibfield
  {journal} {\bibinfo  {journal} {Phys. Rev. Lett.}\ }\textbf {\bibinfo
  {volume} {103}},\ \bibinfo {pages} {236401} (\bibinfo {year}
  {2009})}\BibitemShut {NoStop}%
\bibitem [{\citenamefont {Bhoi}\ \emph {et~al.}(2016)\citenamefont {Bhoi},
  \citenamefont {Khim}, \citenamefont {Nam}, \citenamefont {Lee}, \citenamefont
  {Kim}, \citenamefont {Jeon}, \citenamefont {Min}, \citenamefont {Park},\ and\
  \citenamefont {Kim}}]{bhoi_interplay_2016}%
  \BibitemOpen
  \bibfield  {author} {\bibinfo {author} {\bibfnamefont {D.}~\bibnamefont
  {Bhoi}}, \bibinfo {author} {\bibfnamefont {S.}~\bibnamefont {Khim}}, \bibinfo
  {author} {\bibfnamefont {W.}~\bibnamefont {Nam}}, \bibinfo {author}
  {\bibfnamefont {B.~S.}\ \bibnamefont {Lee}}, \bibinfo {author} {\bibfnamefont
  {C.}~\bibnamefont {Kim}}, \bibinfo {author} {\bibfnamefont {B.-G.}\
  \bibnamefont {Jeon}}, \bibinfo {author} {\bibfnamefont {B.~H.}\ \bibnamefont
  {Min}}, \bibinfo {author} {\bibfnamefont {S.}~\bibnamefont {Park}}, \ and\
  \bibinfo {author} {\bibfnamefont {K.~H.}\ \bibnamefont {Kim}},\ }\href
  {\doibase 10.1038/srep24068} {\bibfield  {journal} {\bibinfo  {journal} {Sci.
  Rep.}\ }\textbf {\bibinfo {volume} {6}},\ \bibinfo {pages} {24068} (\bibinfo
  {year} {2016})}\BibitemShut {NoStop}%
\bibitem [{\citenamefont {Freitas}\ \emph {et~al.}(2016)\citenamefont
  {Freitas}, \citenamefont {Rodi\`ere}, \citenamefont {Osorio}, \citenamefont
  {Navarro-Moratalla}, \citenamefont {Nemes}, \citenamefont {Tissen},
  \citenamefont {Cario}, \citenamefont {Coronado}, \citenamefont
  {Garc\'{\i}a-Hern\'andez}, \citenamefont {Vieira}, \citenamefont {N\'u\~nez
  Regueiro},\ and\ \citenamefont {Suderow}}]{freitas_strong_2016}%
  \BibitemOpen
  \bibfield  {author} {\bibinfo {author} {\bibfnamefont {D.~C.}\ \bibnamefont
  {Freitas}}, \bibinfo {author} {\bibfnamefont {P.}~\bibnamefont {Rodi\`ere}},
  \bibinfo {author} {\bibfnamefont {M.~R.}\ \bibnamefont {Osorio}}, \bibinfo
  {author} {\bibfnamefont {E.}~\bibnamefont {Navarro-Moratalla}}, \bibinfo
  {author} {\bibfnamefont {N.~M.}\ \bibnamefont {Nemes}}, \bibinfo {author}
  {\bibfnamefont {V.~G.}\ \bibnamefont {Tissen}}, \bibinfo {author}
  {\bibfnamefont {L.}~\bibnamefont {Cario}}, \bibinfo {author} {\bibfnamefont
  {E.}~\bibnamefont {Coronado}}, \bibinfo {author} {\bibfnamefont
  {M.}~\bibnamefont {Garc\'{\i}a-Hern\'andez}}, \bibinfo {author}
  {\bibfnamefont {S.}~\bibnamefont {Vieira}}, \bibinfo {author} {\bibfnamefont
  {M.}~\bibnamefont {N\'u\~nez Regueiro}}, \ and\ \bibinfo {author}
  {\bibfnamefont {H.}~\bibnamefont {Suderow}},\ }\href {\doibase
  10.1103/PhysRevB.93.184512} {\bibfield  {journal} {\bibinfo  {journal} {Phys.
  Rev. B}\ }\textbf {\bibinfo {volume} {93}},\ \bibinfo {pages} {184512}
  (\bibinfo {year} {2016})}\BibitemShut {NoStop}%
\bibitem [{\citenamefont {Navarro-Moratalla}\ \emph {et~al.}(2016)\citenamefont
  {Navarro-Moratalla}, \citenamefont {Island}, \citenamefont {Ma\~nas Valero},
  \citenamefont {Pinilla-Cienfuegos}, \citenamefont {Castellanos-Gomez},
  \citenamefont {Quereda}, \citenamefont {Rubio-Bollinger}, \citenamefont
  {Chirolli}, \citenamefont {Silva-Guill\'en}, \citenamefont {Agraït},
  \citenamefont {Steele}, \citenamefont {Guinea}, \citenamefont {Zant},\ and\
  \citenamefont {Coronado}}]{navarro-moratalla_enhanced_2016}%
  \BibitemOpen
  \bibfield  {author} {\bibinfo {author} {\bibfnamefont {E.}~\bibnamefont
  {Navarro-Moratalla}}, \bibinfo {author} {\bibfnamefont {J.~O.}\ \bibnamefont
  {Island}}, \bibinfo {author} {\bibfnamefont {S.}~\bibnamefont {Ma\~nas
  Valero}}, \bibinfo {author} {\bibfnamefont {E.}~\bibnamefont
  {Pinilla-Cienfuegos}}, \bibinfo {author} {\bibfnamefont {A.}~\bibnamefont
  {Castellanos-Gomez}}, \bibinfo {author} {\bibfnamefont {J.}~\bibnamefont
  {Quereda}}, \bibinfo {author} {\bibfnamefont {G.}~\bibnamefont
  {Rubio-Bollinger}}, \bibinfo {author} {\bibfnamefont {L.}~\bibnamefont
  {Chirolli}}, \bibinfo {author} {\bibfnamefont {J.~A.}\ \bibnamefont
  {Silva-Guill\'en}}, \bibinfo {author} {\bibfnamefont {N.}~\bibnamefont
  {Agraït}}, \bibinfo {author} {\bibfnamefont {G.~A.}\ \bibnamefont {Steele}},
  \bibinfo {author} {\bibfnamefont {F.}~\bibnamefont {Guinea}}, \bibinfo
  {author} {\bibfnamefont {H.~S. J. v.~d.}\ \bibnamefont {Zant}}, \ and\
  \bibinfo {author} {\bibfnamefont {E.}~\bibnamefont {Coronado}},\ }\href
  {\doibase 10.1038/ncomms11043} {\bibfield  {journal} {\bibinfo  {journal}
  {Nat. Commun.}\ }\textbf {\bibinfo {volume} {7}},\ \bibinfo {pages} {11043}
  (\bibinfo {year} {2016})}\BibitemShut {NoStop}%
\bibitem [{\citenamefont {Buhot}\ \emph {et~al.}(2015)\citenamefont {Buhot},
  \citenamefont {Toulouse}, \citenamefont {Gallais}, \citenamefont {Sacuto},
  \citenamefont {de~Sousa}, \citenamefont {Wang}, \citenamefont {Bellaiche},
  \citenamefont {Bibes}, \citenamefont {Barth\'el\'emy}, \citenamefont
  {Forget}, \citenamefont {Colson}, \citenamefont {Cazayous},\ and\
  \citenamefont {M\'easson}}]{buhot_driving_2015}%
  \BibitemOpen
  \bibfield  {author} {\bibinfo {author} {\bibfnamefont {J.}~\bibnamefont
  {Buhot}}, \bibinfo {author} {\bibfnamefont {C.}~\bibnamefont {Toulouse}},
  \bibinfo {author} {\bibfnamefont {Y.}~\bibnamefont {Gallais}}, \bibinfo
  {author} {\bibfnamefont {A.}~\bibnamefont {Sacuto}}, \bibinfo {author}
  {\bibfnamefont {R.}~\bibnamefont {de~Sousa}}, \bibinfo {author}
  {\bibfnamefont {D.}~\bibnamefont {Wang}}, \bibinfo {author} {\bibfnamefont
  {L.}~\bibnamefont {Bellaiche}}, \bibinfo {author} {\bibfnamefont
  {M.}~\bibnamefont {Bibes}}, \bibinfo {author} {\bibfnamefont
  {A.}~\bibnamefont {Barth\'el\'emy}}, \bibinfo {author} {\bibfnamefont
  {A.}~\bibnamefont {Forget}}, \bibinfo {author} {\bibfnamefont
  {D.}~\bibnamefont {Colson}}, \bibinfo {author} {\bibfnamefont
  {M.}~\bibnamefont {Cazayous}}, \ and\ \bibinfo {author} {\bibfnamefont
  {M.-A.}\ \bibnamefont {M\'easson}},\ }\href {\doibase
  10.1103/PhysRevLett.115.267204} {\bibfield  {journal} {\bibinfo  {journal}
  {Phys. Rev. Lett.}\ }\textbf {\bibinfo {volume} {115}},\ \bibinfo {pages}
  {267204} (\bibinfo {year} {2015})}\BibitemShut {NoStop}%
\bibitem [{\citenamefont {Sugai}\ \emph {et~al.}(1981)\citenamefont {Sugai},
  \citenamefont {Murase}, \citenamefont {Uchida},\ and\ \citenamefont
  {Tanaka}}]{sugai_studies_1981}%
  \BibitemOpen
  \bibfield  {author} {\bibinfo {author} {\bibfnamefont {S.}~\bibnamefont
  {Sugai}}, \bibinfo {author} {\bibfnamefont {K.}~\bibnamefont {Murase}},
  \bibinfo {author} {\bibfnamefont {S.}~\bibnamefont {Uchida}}, \ and\ \bibinfo
  {author} {\bibfnamefont {S.}~\bibnamefont {Tanaka}},\ }\href {\doibase
  10.1016/0038-1098(81)90847-4} {\bibfield  {journal} {\bibinfo  {journal}
  {Solid State Commun.}\ }\textbf {\bibinfo {volume} {40}},\ \bibinfo {pages}
  {399} (\bibinfo {year} {1981})}\BibitemShut {NoStop}%
\bibitem [{Note1()}]{Note1}%
  \BibitemOpen
  \bibinfo {note} {$(1-{\begingroup x^4\endgroup \over 3})\protect \sqrt
  {1-x^4}$ where $x={\begingroup T\endgroup \over T_{CDW}}$ taken from
  Ref.\cite {Benfatto2000} and which reproduces a mean-field-like behaviour
  near both $T=0$ and $T_{CDW}$}\BibitemShut {NoStop}%
\bibitem [{\citenamefont {Eiter}\ \emph {et~al.}(2013)\citenamefont {Eiter},
  \citenamefont {Lavagnini}, \citenamefont {Hackl}, \citenamefont {Nowadnick},
  \citenamefont {Kemper}, \citenamefont {Devereaux}, \citenamefont {Chu},
  \citenamefont {Analytis}, \citenamefont {Fisher},\ and\ \citenamefont
  {Degiorgi}}]{eiter_alternative_2013}%
  \BibitemOpen
  \bibfield  {author} {\bibinfo {author} {\bibfnamefont {H.-M.}\ \bibnamefont
  {Eiter}}, \bibinfo {author} {\bibfnamefont {M.}~\bibnamefont {Lavagnini}},
  \bibinfo {author} {\bibfnamefont {R.}~\bibnamefont {Hackl}}, \bibinfo
  {author} {\bibfnamefont {E.~A.}\ \bibnamefont {Nowadnick}}, \bibinfo {author}
  {\bibfnamefont {A.~F.}\ \bibnamefont {Kemper}}, \bibinfo {author}
  {\bibfnamefont {T.~P.}\ \bibnamefont {Devereaux}}, \bibinfo {author}
  {\bibfnamefont {J.-H.}\ \bibnamefont {Chu}}, \bibinfo {author} {\bibfnamefont
  {J.~G.}\ \bibnamefont {Analytis}}, \bibinfo {author} {\bibfnamefont {I.~R.}\
  \bibnamefont {Fisher}}, \ and\ \bibinfo {author} {\bibfnamefont
  {L.}~\bibnamefont {Degiorgi}},\ }\href {\doibase 10.1073/pnas.1214745110}
  {\bibfield  {journal} {\bibinfo  {journal} {Proc. Natl. Acad. Sci.}\ }\textbf
  {\bibinfo {volume} {110}},\ \bibinfo {pages} {64} (\bibinfo {year}
  {2013})}\BibitemShut {NoStop}%
\bibitem [{\citenamefont {Ralevi\ifmmode~\acute{c}\else \'{c}\fi{}}\ \emph
  {et~al.}(2016)\citenamefont {Ralevi\ifmmode~\acute{c}\else \'{c}\fi{}},
  \citenamefont {Lazarevi\ifmmode~\acute{c}\else \'{c}\fi{}}, \citenamefont
  {Baum}, \citenamefont {Eiter}, \citenamefont {Hackl}, \citenamefont
  {Giraldo-Gallo}, \citenamefont {Fisher}, \citenamefont {Petrovic},
  \citenamefont {Gaji\ifmmode~\acute{c}\else \'{c}\fi{}},\ and\ \citenamefont
  {Popovi\ifmmode~\acute{c}\else \'{c}\fi{}}}]{ralevic_charge_2016}%
  \BibitemOpen
  \bibfield  {author} {\bibinfo {author} {\bibfnamefont {U.}~\bibnamefont
  {Ralevi\ifmmode~\acute{c}\else \'{c}\fi{}}}, \bibinfo {author} {\bibfnamefont
  {N.}~\bibnamefont {Lazarevi\ifmmode~\acute{c}\else \'{c}\fi{}}}, \bibinfo
  {author} {\bibfnamefont {A.}~\bibnamefont {Baum}}, \bibinfo {author}
  {\bibfnamefont {H.-M.}\ \bibnamefont {Eiter}}, \bibinfo {author}
  {\bibfnamefont {R.}~\bibnamefont {Hackl}}, \bibinfo {author} {\bibfnamefont
  {P.}~\bibnamefont {Giraldo-Gallo}}, \bibinfo {author} {\bibfnamefont {I.~R.}\
  \bibnamefont {Fisher}}, \bibinfo {author} {\bibfnamefont {C.}~\bibnamefont
  {Petrovic}}, \bibinfo {author} {\bibfnamefont {R.}~\bibnamefont
  {Gaji\ifmmode~\acute{c}\else \'{c}\fi{}}}, \ and\ \bibinfo {author}
  {\bibfnamefont {Z.~V.}\ \bibnamefont {Popovi\ifmmode~\acute{c}\else
  \'{c}\fi{}}},\ }\href {\doibase 10.1103/PhysRevB.94.165132} {\bibfield
  {journal} {\bibinfo  {journal} {Phys. Rev. B}\ }\textbf {\bibinfo {volume}
  {94}},\ \bibinfo {pages} {165132} (\bibinfo {year} {2016})}\BibitemShut
  {NoStop}%
\bibitem [{\citenamefont {Zhao}\ \emph {et~al.}(2017)\citenamefont {Zhao},
  \citenamefont {Wijayaratne}, \citenamefont {Butler}, \citenamefont {Yang},
  \citenamefont {Malliakas}, \citenamefont {Chung}, \citenamefont {Louca},
  \citenamefont {Kanatzidis}, \citenamefont {van Wezel},\ and\ \citenamefont
  {Chatterjee}}]{zhao_orbital_2017}%
  \BibitemOpen
  \bibfield  {author} {\bibinfo {author} {\bibfnamefont {J.}~\bibnamefont
  {Zhao}}, \bibinfo {author} {\bibfnamefont {K.}~\bibnamefont {Wijayaratne}},
  \bibinfo {author} {\bibfnamefont {A.}~\bibnamefont {Butler}}, \bibinfo
  {author} {\bibfnamefont {J.}~\bibnamefont {Yang}}, \bibinfo {author}
  {\bibfnamefont {C.~D.}\ \bibnamefont {Malliakas}}, \bibinfo {author}
  {\bibfnamefont {D.~Y.}\ \bibnamefont {Chung}}, \bibinfo {author}
  {\bibfnamefont {D.}~\bibnamefont {Louca}}, \bibinfo {author} {\bibfnamefont
  {M.~G.}\ \bibnamefont {Kanatzidis}}, \bibinfo {author} {\bibfnamefont
  {J.}~\bibnamefont {van Wezel}}, \ and\ \bibinfo {author} {\bibfnamefont
  {U.}~\bibnamefont {Chatterjee}},\ }\href {\doibase
  10.1103/PhysRevB.96.125103} {\bibfield  {journal} {\bibinfo  {journal} {Phys.
  Rev. B}\ }\textbf {\bibinfo {volume} {96}},\ \bibinfo {pages} {125103}
  (\bibinfo {year} {2017})}\BibitemShut {NoStop}%
\bibitem [{Note2()}]{Note2}%
  \BibitemOpen
  \bibinfo {note} {Equation of Ref.[39] using $x={\begingroup P\endgroup \over
  P_c}$}\BibitemShut {NoStop}%
\bibitem [{\citenamefont {Chen}\ \emph {et~al.}(1997)\citenamefont {Chen},
  \citenamefont {Naeini}, \citenamefont {Hewitt}, \citenamefont {Irwin},
  \citenamefont {Liang},\ and\ \citenamefont {Hardy}}]{chen_electronic_1997}%
  \BibitemOpen
  \bibfield  {author} {\bibinfo {author} {\bibfnamefont {X.~K.}\ \bibnamefont
  {Chen}}, \bibinfo {author} {\bibfnamefont {J.~G.}\ \bibnamefont {Naeini}},
  \bibinfo {author} {\bibfnamefont {K.~C.}\ \bibnamefont {Hewitt}}, \bibinfo
  {author} {\bibfnamefont {J.~C.}\ \bibnamefont {Irwin}}, \bibinfo {author}
  {\bibfnamefont {R.}~\bibnamefont {Liang}}, \ and\ \bibinfo {author}
  {\bibfnamefont {W.~N.}\ \bibnamefont {Hardy}},\ }\href {\doibase
  10.1103/PhysRevB.56.R513} {\bibfield  {journal} {\bibinfo  {journal} {Phys.
  Rev. B}\ }\textbf {\bibinfo {volume} {56}},\ \bibinfo {pages} {R513}
  (\bibinfo {year} {1997})}\BibitemShut {NoStop}%
\bibitem [{\citenamefont {Sugai}\ \emph {et~al.}(2003)\citenamefont {Sugai},
  \citenamefont {Suzuki}, \citenamefont {Takayanagi}, \citenamefont
  {Hosokawa},\ and\ \citenamefont
  {Hayamizu}}]{sugai_carrier-density-dependent_2003}%
  \BibitemOpen
  \bibfield  {author} {\bibinfo {author} {\bibfnamefont {S.}~\bibnamefont
  {Sugai}}, \bibinfo {author} {\bibfnamefont {H.}~\bibnamefont {Suzuki}},
  \bibinfo {author} {\bibfnamefont {Y.}~\bibnamefont {Takayanagi}}, \bibinfo
  {author} {\bibfnamefont {T.}~\bibnamefont {Hosokawa}}, \ and\ \bibinfo
  {author} {\bibfnamefont {N.}~\bibnamefont {Hayamizu}},\ }\href {\doibase
  10.1103/PhysRevB.68.184504} {\bibfield  {journal} {\bibinfo  {journal} {Phys.
  Rev. B}\ }\textbf {\bibinfo {volume} {68}},\ \bibinfo {pages} {184504}
  (\bibinfo {year} {2003})}\BibitemShut {NoStop}%
\bibitem [{\citenamefont {Klein}\ and\ \citenamefont
  {Dierker}(1984)}]{klein_1984}%
  \BibitemOpen
  \bibfield  {author} {\bibinfo {author} {\bibfnamefont {M.~V.}\ \bibnamefont
  {Klein}}\ and\ \bibinfo {author} {\bibfnamefont {S.~B.}\ \bibnamefont
  {Dierker}},\ }\href {\doibase 10.1103/PhysRevB.29.4976} {\bibfield  {journal}
  {\bibinfo  {journal} {Phys. Rev. B}\ }\textbf {\bibinfo {volume} {29}},\
  \bibinfo {pages} {4976} (\bibinfo {year} {1984})}\BibitemShut {NoStop}%
\bibitem [{\citenamefont {Devereaux}\ and\ \citenamefont
  {Einzel}(1995)}]{devereaux_electronic_1995}%
  \BibitemOpen
  \bibfield  {author} {\bibinfo {author} {\bibfnamefont {T.~P.}\ \bibnamefont
  {Devereaux}}\ and\ \bibinfo {author} {\bibfnamefont {D.}~\bibnamefont
  {Einzel}},\ }\href {\doibase 10.1103/PhysRevB.51.16336} {\bibfield  {journal}
  {\bibinfo  {journal} {Phys. Rev. B}\ }\textbf {\bibinfo {volume} {51}},\
  \bibinfo {pages} {16336} (\bibinfo {year} {1995})}\BibitemShut {NoStop}%
\bibitem [{\citenamefont {Devereaux}\ and\ \citenamefont
  {Hackl}(2007)}]{devereaux_inelastic_2007}%
  \BibitemOpen
  \bibfield  {author} {\bibinfo {author} {\bibfnamefont {T.~P.}\ \bibnamefont
  {Devereaux}}\ and\ \bibinfo {author} {\bibfnamefont {R.}~\bibnamefont
  {Hackl}},\ }\href {\doibase 10.1103/RevModPhys.79.175} {\bibfield  {journal}
  {\bibinfo  {journal} {Rev. Mod. Phys.}\ }\textbf {\bibinfo {volume} {79}},\
  \bibinfo {pages} {175} (\bibinfo {year} {2007})}\BibitemShut {NoStop}%
\bibitem [{\citenamefont {Cea}\ and\ \citenamefont
  {Benfatto}(2016)}]{cea_raman_prb16}%
  \BibitemOpen
  \bibfield  {author} {\bibinfo {author} {\bibfnamefont {T.}~\bibnamefont
  {Cea}}\ and\ \bibinfo {author} {\bibfnamefont {L.}~\bibnamefont {Benfatto}},\
  }\href {\doibase 10.1103/PhysRevB.94.064512} {\bibfield  {journal} {\bibinfo
  {journal} {Phys. Rev. B}\ }\textbf {\bibinfo {volume} {94}},\ \bibinfo
  {pages} {064512} (\bibinfo {year} {2016})}\BibitemShut {NoStop}%
\bibitem [{\citenamefont {Benfatto}\ \emph {et~al.}(2000)\citenamefont
  {Benfatto}, \citenamefont {Caprara},\ and\ \citenamefont
  {Castro}}]{Benfatto2000}%
  \BibitemOpen
  \bibfield  {author} {\bibinfo {author} {\bibfnamefont {L.}~\bibnamefont
  {Benfatto}}, \bibinfo {author} {\bibfnamefont {S.}~\bibnamefont {Caprara}}, \
  and\ \bibinfo {author} {\bibfnamefont {C.~D.}\ \bibnamefont {Castro}},\
  }\href {\doibase 10.1007/s100510070163} {\bibfield  {journal} {\bibinfo
  {journal} {Eur. Phys. J. B}\ }\textbf {\bibinfo {volume} {17}},\ \bibinfo
  {pages} {95} (\bibinfo {year} {2000})}\BibitemShut {NoStop}%
\end{thebibliography}

%





%

\end{document}